\RequirePackage{lineno}
\documentclass[aps,prl,superscriptaddress,groupedaddress,preprintnumbers,nofootinbib]{revtex4}  
\usepackage{graphicx}  
\usepackage{dcolumn}   
\usepackage{bm}        
\usepackage{amssymb}   
\usepackage{amsmath}
\usepackage{hyperref}

\usepackage[normalem]{ulem}
\usepackage{xcolor}

\newcommand{\de}{\mathrm d}

\newcommand{\g}{$\gamma$}
\newcommand{\bi}{\begin{itemize}}
\newcommand{\ei}{\end{itemize}}
\newcommand{\be}{\begin{equation}}
\newcommand{\ee}{\end{equation}}
\newcommand{\nn}{\mathcal{N}}
\newcommand{\Fermi}{\textit{Fermi}-LAT}

\newcommand{\sv}{\langle\sigma_{\rm ann} v\rangle}
\newcommand{\svth}{\langle\sigma_{\rm ann} v\rangle_{\rm th}}
\newcommand{\mdm}{m_{\rm DM}}
\newcommand{\odm}{\Omega_\mathrm{DM}}
\newcommand{\ob}{\Omega_\mathrm{b}}
\newcommand{\om}{\Omega_{\rm m}}
\newcommand{\DM}{dark matter}

\newcommand{\phys}{\textit{physical}}
\newcommand{\pheno}{\textit{phenomenological}}
\newcommand{\deltachi}{$\Delta\chi^2$}

\begin{document}

\title{Detection of cross-correlation between gravitational lensing and gamma rays}
\affiliation{Dipartimento di Fisica, Universit\`a degli Studi di Torino, via P. Giuria 1, 10125 Torino, Italy}
\affiliation{INFN -- Istituto Nazionale di Fisica Nucleare, Sezione di Torino, via P. Giuria 1, 10125 Torino, Italy}
\affiliation{Kavli Institute for Particle Astrophysics \& Cosmology, P. O. Box 2450, Stanford University, Stanford, CA 94305, USA}
\affiliation{SLAC National Accelerator Laboratory, Menlo Park, CA 94025, USA}
\affiliation{INAF -- Istituto Nazionale di Astrofisica, Osservatorio Astrofisico di Torino, strada Osservatorio 20, 10025 Pino Torinese, Italy}
\affiliation{Department of Physics \& Astronomy, University of the Western Cape, Cape Town 7535, South Africa}
\affiliation{GRAPPA Institute, University of Amsterdam, 1098 XH Amsterdam, The Netherlands}
\affiliation{Kavli Institute for the Physics and Mathematics of the Universe, University of Tokyo, Kashiwa, Chiba 277-8583, Japan}
\affiliation{LSST, 933 North Cherry Avenue, Tucson, AZ 85721, USA}
\affiliation{Physics Department, 2320 Chamberlin Hall, University of Wisconsin-Madison, 1150 University Avenue Madison, WI  53706-1390}
\affiliation{Jodrell Bank Center for Astrophysics, School of Physics and Astronomy, University of Manchester, Oxford Road, Manchester, M13 9PL, UK}
\affiliation{Center for Cosmology and Astro-Particle Physics, The Ohio State University, Columbus, OH 43210, USA}
\affiliation{Department of Astronomy/Steward Observatory, University of Arizona, 933 North Cherry Avenue, Tucson, AZ 85721-0065, USA}
\affiliation{Jet Propulsion Laboratory, California Institute of Technology, 4800 Oak Grove Dr., Pasadena, CA 91109, USA}
\affiliation{Institut de F\'{\i}sica d'Altes Energies (IFAE), The Barcelona Institute of Science and Technology, Campus UAB, 08193 Bellaterra (Barcelona) Spain}
\affiliation{Center for Cosmology and Astro-Particle Physics, The Ohio State University, Columbus, OH 43210, USA}
\affiliation{Department of Physics, The Ohio State University, Columbus, OH 43210, USA}
\affiliation{Department of Physics, Carnegie Mellon University, Pittsburgh, Pennsylvania 15312, USA}
\affiliation{Brookhaven National Laboratory, Bldg 510, Upton, NY 11973, USA}
\affiliation{Department of Physics, Duke University Durham, NC 27708, USA}
\affiliation{Institute for Astronomy, University of Edinburgh, Edinburgh EH9 3HJ, UK}
\affiliation{Department of Astronomy, University of Illinois at Urbana-Champaign, 1002 W. Green Street, Urbana, IL 61801, USA}
\affiliation{National Center for Supercomputing Applications, 1205 West Clark St., Urbana, IL 61801, USA}
\affiliation{Fermi National Accelerator Laboratory, P. O. Box 500, Batavia, IL 60510, USA}
\affiliation{Instituto de Fisica Teorica UAM/CSIC, Universidad Autonoma de Madrid, 28049 Madrid, Spain}
\affiliation{CNRS, UMR 7095, Institut d'Astrophysique de Paris, F-75014, Paris, France}
\affiliation{Sorbonne Universit\'es, UPMC Univ Paris 06, UMR 7095, Institut d'Astrophysique de Paris, F-75014, Paris, France}
\affiliation{Department of Physics \& Astronomy, University College London, Gower Street, London, WC1E 6BT, UK}
\affiliation{Centro de Investigaciones Energ\'eticas, Medioambientales y Tecnol\'ogicas (CIEMAT), Madrid, Spain}
\affiliation{Laborat\'orio Interinstitucional de e-Astronomia - LIneA, Rua Gal. Jos\'e Cristino 77, Rio de Janeiro, RJ - 20921-400, Brazil}
\affiliation{Institut d'Estudis Espacials de Catalunya (IEEC), 08034 Barcelona, Spain}
\affiliation{Institute of Space Sciences (ICE, CSIC),  Campus UAB, Carrer de Can Magrans, s/n,  08193 Barcelona, Spain}
\affiliation{INAF -- Osservatorio Astronomico di Trieste, via G.~B.~Tiepolo 11, 34143 Trieste, Italy}
\affiliation{IFPU -- Institute for Fundamental Physics of the Universe, Via Beirut 2, 34014 Trieste, Italy}
\affiliation{Observat\'orio Nacional, Rua Gal. Jos\'e Cristino 77, Rio de Janeiro, RJ - 20921-400, Brazil}
\affiliation{Department of Physics, IIT Hyderabad, Kandi, Telangana 502285, India}
\affiliation{Excellence Cluster Origins, Boltzmannstr.\ 2, 85748 Garching, Germany}
\affiliation{Faculty of Physics, Ludwig-Maximilians-Universit\"at, Scheinerstr. 1, 81679 Munich, Germany}
\affiliation{Santa Cruz Institute for Particle Physics, Santa Cruz, CA 95064, USA}
\affiliation{Department of Astronomy, University of Michigan, Ann Arbor, MI 48109, USA}
\affiliation{Department of Physics, University of Michigan, Ann Arbor, MI 48109, USA}
\affiliation{Institute of Astronomy, University of Cambridge, Madingley Road, Cambridge CB3 0HA, UK}
\affiliation{Kavli Institute for Cosmology, University of Cambridge, Madingley Road, Cambridge CB3 0HA, UK}
\affiliation{California Institute of Technology, 1200 East California Blvd, MC 249-17, Pasadena, CA 91125, USA}
\affiliation{Center for Astrophysics $\vert$ Harvard \& Smithsonian, 60 Garden Street, Cambridge, MA 02138, USA}
\affiliation{Department of Physics and Astronomy, University of Pennsylvania, Philadelphia, PA 19104, USA}
\affiliation{Kavli Institute for Cosmological Physics, University of Chicago, Chicago, IL 60637, USA}
\affiliation{Departamento de F\'isica Matem\'atica, Instituto de F\'isica, Universidade de S\~ao Paulo, CP 66318, S\~ao Paulo, SP, 05314-970, Brazil}
\affiliation{George P. and Cynthia Woods Mitchell Institute for Fundamental Physics and Astronomy, and Department of Physics and Astronomy, Texas A\&M University, College Station, TX 77843,  USA}
\affiliation{Department of Astrophysical Sciences, Princeton University, Peyton Hall, Princeton, NJ 08544, USA}
\affiliation{Instituci\'o Catalana de Recerca i Estudis Avan\c{c}ats, E-08010 Barcelona, Spain}
\affiliation{School of Physics and Astronomy, University of Southampton,  Southampton, SO17 1BJ, UK}
\affiliation{Brandeis University, Physics Department, 415 South Street, Waltham MA 02453}
\affiliation{Instituto de F\'isica Gleb Wataghin, Universidade Estadual de Campinas, 13083-859, Campinas, SP, Brazil}
\affiliation{Computer Science and Mathematics Division, Oak Ridge National Laboratory, Oak Ridge, TN 37831}
\affiliation{Institute of Cosmology and Gravitation, University of Portsmouth, Portsmouth, PO1 3FX, UK}
\affiliation{Argonne National Laboratory, 9700 South Cass Avenue, Lemont, IL 60439, USA}

\author{S.~Ammazzalorso\footnote{simone.ammazzalorso@unito.it}}
\affiliation{Dipartimento di Fisica, Universit\`a degli Studi di Torino, via P. Giuria 1, 10125 Torino, Italy}
\affiliation{INFN -- Istituto Nazionale di Fisica Nucleare, Sezione di Torino, via P. Giuria 1, 10125 Torino, Italy}
\author{D.~Gruen\footnote{dgruen@stanford.edu}}
\affiliation{Kavli Institute for Particle Astrophysics \& Cosmology, P. O. Box 2450, Stanford University, Stanford, CA 94305, USA}
\affiliation{SLAC National Accelerator Laboratory, Menlo Park, CA 94025, USA}
\author{M.~Regis}
\affiliation{Dipartimento di Fisica, Universit\`a degli Studi di Torino, via P. Giuria 1, 10125 Torino, Italy}
\affiliation{INFN -- Istituto Nazionale di Fisica Nucleare, Sezione di Torino, via P. Giuria 1, 10125 Torino, Italy}
\author{S.~Camera}
\affiliation{Dipartimento di Fisica, Universit\`a degli Studi di Torino, via P. Giuria 1, 10125 Torino, Italy}
\affiliation{INFN -- Istituto Nazionale di Fisica Nucleare, Sezione di Torino, via P. Giuria 1, 10125 Torino, Italy}
\affiliation{INAF -- Istituto Nazionale di Astrofisica, Osservatorio Astrofisico di Torino, strada Osservatorio 20, 10025 Pino Torinese, Italy}
\affiliation{Department of Physics \& Astronomy, University of the Western Cape, Cape Town 7535, South Africa}
\author{S.~Ando}
\affiliation{GRAPPA Institute, University of Amsterdam, 1098 XH Amsterdam, The Netherlands}
\affiliation{Kavli Institute for the Physics and Mathematics of the Universe, University of Tokyo, Kashiwa, Chiba 277-8583, Japan}
\author{N.~Fornengo}
\affiliation{Dipartimento di Fisica, Universit\`a degli Studi di Torino, via P. Giuria 1, 10125 Torino, Italy}
\affiliation{INFN -- Istituto Nazionale di Fisica Nucleare, Sezione di Torino, via P. Giuria 1, 10125 Torino, Italy}
\author{K.~Bechtol}
\affiliation{LSST, 933 North Cherry Avenue, Tucson, AZ 85721, USA}
\affiliation{Physics Department, 2320 Chamberlin Hall, University of Wisconsin-Madison, 1150 University Avenue Madison, WI  53706-1390}
\author{S.~L.~Bridle}
\affiliation{Jodrell Bank Center for Astrophysics, School of Physics and Astronomy, University of Manchester, Oxford Road, Manchester, M13 9PL, UK}
\author{A.~Choi}
\affiliation{Center for Cosmology and Astro-Particle Physics, The Ohio State University, Columbus, OH 43210, USA}
\author{T.~F.~Eifler}
\affiliation{Department of Astronomy/Steward Observatory, University of Arizona, 933 North Cherry Avenue, Tucson, AZ 85721-0065, USA}
\affiliation{Jet Propulsion Laboratory, California Institute of Technology, 4800 Oak Grove Dr., Pasadena, CA 91109, USA}
\author{M.~Gatti}
\affiliation{Institut de F\'{\i}sica d'Altes Energies (IFAE), The Barcelona Institute of Science and Technology, Campus UAB, 08193 Bellaterra (Barcelona) Spain}
\author{N.~MacCrann}
\affiliation{Center for Cosmology and Astro-Particle Physics, The Ohio State University, Columbus, OH 43210, USA}
\affiliation{Department of Physics, The Ohio State University, Columbus, OH 43210, USA}
\author{Y.~Omori}
\affiliation{Kavli Institute for Particle Astrophysics \& Cosmology, P. O. Box 2450, Stanford University, Stanford, CA 94305, USA}
\author{S.~Samuroff}
\affiliation{Department of Physics, Carnegie Mellon University, Pittsburgh, Pennsylvania 15312, USA}
\author{E.~Sheldon}
\affiliation{Brookhaven National Laboratory, Bldg 510, Upton, NY 11973, USA}
\author{M.~A.~Troxel}
\affiliation{Department of Physics, Duke University Durham, NC 27708, USA}
\author{J.~Zuntz}
\affiliation{Institute for Astronomy, University of Edinburgh, Edinburgh EH9 3HJ, UK}
\author{M.~Carrasco~Kind}
\affiliation{Department of Astronomy, University of Illinois at Urbana-Champaign, 1002 W. Green Street, Urbana, IL 61801, USA}
\affiliation{National Center for Supercomputing Applications, 1205 West Clark St., Urbana, IL 61801, USA}
\author{J.~Annis}
\affiliation{Fermi National Accelerator Laboratory, P. O. Box 500, Batavia, IL 60510, USA}
\author{S.~Avila}
\affiliation{Instituto de Fisica Teorica UAM/CSIC, Universidad Autonoma de Madrid, 28049 Madrid, Spain}
\author{E.~Bertin}
\affiliation{CNRS, UMR 7095, Institut d'Astrophysique de Paris, F-75014, Paris, France}
\affiliation{Sorbonne Universit\'es, UPMC Univ Paris 06, UMR 7095, Institut d'Astrophysique de Paris, F-75014, Paris, France}
\author{D.~Brooks}
\affiliation{Department of Physics \& Astronomy, University College London, Gower Street, London, WC1E 6BT, UK}
\author{D.~L.~Burke}
\affiliation{Kavli Institute for Particle Astrophysics \& Cosmology, P. O. Box 2450, Stanford University, Stanford, CA 94305, USA}
\affiliation{SLAC National Accelerator Laboratory, Menlo Park, CA 94025, USA}
\author{A.~Carnero~Rosell}
\affiliation{Centro de Investigaciones Energ\'eticas, Medioambientales y Tecnol\'ogicas (CIEMAT), Madrid, Spain}
\affiliation{Laborat\'orio Interinstitucional de e-Astronomia - LIneA, Rua Gal. Jos\'e Cristino 77, Rio de Janeiro, RJ - 20921-400, Brazil}
\author{J.~Carretero}
\affiliation{Institut de F\'{\i}sica d'Altes Energies (IFAE), The Barcelona Institute of Science and Technology, Campus UAB, 08193 Bellaterra (Barcelona) Spain}
\author{F.~J.~Castander}
\affiliation{Institut d'Estudis Espacials de Catalunya (IEEC), 08034 Barcelona, Spain}
\affiliation{Institute of Space Sciences (ICE, CSIC),  Campus UAB, Carrer de Can Magrans, s/n,  08193 Barcelona, Spain}
\author{M.~Costanzi}
\affiliation{INAF -- Osservatorio Astronomico di Trieste, via G.~B.~Tiepolo 11, 34143 Trieste, Italy}
\affiliation{IFPU -- Institute for Fundamental Physics of the Universe, Via Beirut 2, 34014 Trieste, Italy}
\author{L.~N.~da Costa}
\affiliation{Laborat\'orio Interinstitucional de e-Astronomia - LIneA, Rua Gal. Jos\'e Cristino 77, Rio de Janeiro, RJ - 20921-400, Brazil}
\affiliation{Observat\'orio Nacional, Rua Gal. Jos\'e Cristino 77, Rio de Janeiro, RJ - 20921-400, Brazil}
\author{J.~De~Vicente}
\affiliation{Centro de Investigaciones Energ\'eticas, Medioambientales y Tecnol\'ogicas (CIEMAT), Madrid, Spain}
\author{S.~Desai}
\affiliation{Department of Physics, IIT Hyderabad, Kandi, Telangana 502285, India}
\author{H.~T.~Diehl}
\affiliation{Fermi National Accelerator Laboratory, P. O. Box 500, Batavia, IL 60510, USA}
\author{J.~P.~Dietrich}
\affiliation{Excellence Cluster Origins, Boltzmannstr.\ 2, 85748 Garching, Germany}
\affiliation{Faculty of Physics, Ludwig-Maximilians-Universit\"at, Scheinerstr. 1, 81679 Munich, Germany}
\author{P.~Doel}
\affiliation{Department of Physics \& Astronomy, University College London, Gower Street, London, WC1E 6BT, UK}
\author{S.~Everett}
\affiliation{Santa Cruz Institute for Particle Physics, Santa Cruz, CA 95064, USA}
\author{B.~Flaugher}
\affiliation{Fermi National Accelerator Laboratory, P. O. Box 500, Batavia, IL 60510, USA}
\author{P.~Fosalba}
\affiliation{Institut d'Estudis Espacials de Catalunya (IEEC), 08034 Barcelona, Spain}
\affiliation{Institute of Space Sciences (ICE, CSIC),  Campus UAB, Carrer de Can Magrans, s/n,  08193 Barcelona, Spain}
\author{J.~Garc\'ia-Bellido}
\affiliation{Instituto de Fisica Teorica UAM/CSIC, Universidad Autonoma de Madrid, 28049 Madrid, Spain}
\author{E.~Gaztanaga}
\affiliation{Institut d'Estudis Espacials de Catalunya (IEEC), 08034 Barcelona, Spain}
\affiliation{Institute of Space Sciences (ICE, CSIC),  Campus UAB, Carrer de Can Magrans, s/n,  08193 Barcelona, Spain}
\author{D.~W.~Gerdes}
\affiliation{Department of Astronomy, University of Michigan, Ann Arbor, MI 48109, USA}
\affiliation{Department of Physics, University of Michigan, Ann Arbor, MI 48109, USA}
\author{T.~Giannantonio}
\affiliation{Institute of Astronomy, University of Cambridge, Madingley Road, Cambridge CB3 0HA, UK}
\affiliation{Kavli Institute for Cosmology, University of Cambridge, Madingley Road, Cambridge CB3 0HA, UK}
\author{D.~A.~Goldstein}
\affiliation{California Institute of Technology, 1200 East California Blvd, MC 249-17, Pasadena, CA 91125, USA}
\author{R.~A.~Gruendl}
\affiliation{Department of Astronomy, University of Illinois at Urbana-Champaign, 1002 W. Green Street, Urbana, IL 61801, USA}
\affiliation{National Center for Supercomputing Applications, 1205 West Clark St., Urbana, IL 61801, USA}
\author{G.~Gutierrez}
\affiliation{Fermi National Accelerator Laboratory, P. O. Box 500, Batavia, IL 60510, USA}
\author{D.~L.~Hollowood}
\affiliation{Santa Cruz Institute for Particle Physics, Santa Cruz, CA 95064, USA}
\author{K.~Honscheid}
\affiliation{Center for Cosmology and Astro-Particle Physics, The Ohio State University, Columbus, OH 43210, USA}
\affiliation{Department of Physics, The Ohio State University, Columbus, OH 43210, USA}
\author{D.~J.~James}
\affiliation{Center for Astrophysics $\vert$ Harvard \& Smithsonian, 60 Garden Street, Cambridge, MA 02138, USA}
\author{M.~Jarvis}
\affiliation{Department of Physics and Astronomy, University of Pennsylvania, Philadelphia, PA 19104, USA}
\author{T.~Jeltema}
\affiliation{Santa Cruz Institute for Particle Physics, Santa Cruz, CA 95064, USA}
\author{S.~Kent}
\affiliation{Fermi National Accelerator Laboratory, P. O. Box 500, Batavia, IL 60510, USA}
\affiliation{Kavli Institute for Cosmological Physics, University of Chicago, Chicago, IL 60637, USA}
\author{N.~Kuropatkin}
\affiliation{Fermi National Accelerator Laboratory, P. O. Box 500, Batavia, IL 60510, USA}
\author{O.~Lahav}
\affiliation{Department of Physics \& Astronomy, University College London, Gower Street, London, WC1E 6BT, UK}
\author{T.~S.~Li}
\affiliation{Fermi National Accelerator Laboratory, P. O. Box 500, Batavia, IL 60510, USA}
\affiliation{Kavli Institute for Cosmological Physics, University of Chicago, Chicago, IL 60637, USA}
\author{M.~Lima}
\affiliation{Departamento de F\'isica Matem\'atica, Instituto de F\'isica, Universidade de S\~ao Paulo, CP 66318, S\~ao Paulo, SP, 05314-970, Brazil}
\affiliation{Laborat\'orio Interinstitucional de e-Astronomia - LIneA, Rua Gal. Jos\'e Cristino 77, Rio de Janeiro, RJ - 20921-400, Brazil}
\author{M.~A.~G.~Maia}
\affiliation{Laborat\'orio Interinstitucional de e-Astronomia - LIneA, Rua Gal. Jos\'e Cristino 77, Rio de Janeiro, RJ - 20921-400, Brazil}
\affiliation{Observat\'orio Nacional, Rua Gal. Jos\'e Cristino 77, Rio de Janeiro, RJ - 20921-400, Brazil}
\author{J.~L.~Marshall}
\affiliation{George P. and Cynthia Woods Mitchell Institute for Fundamental Physics and Astronomy, and Department of Physics and Astronomy, Texas A\&M University, College Station, TX 77843,  USA}
\author{P.~Melchior}
\affiliation{Department of Astrophysical Sciences, Princeton University, Peyton Hall, Princeton, NJ 08544, USA}
\author{F.~Menanteau}
\affiliation{Department of Astronomy, University of Illinois at Urbana-Champaign, 1002 W. Green Street, Urbana, IL 61801, USA}
\affiliation{National Center for Supercomputing Applications, 1205 West Clark St., Urbana, IL 61801, USA}
\author{R.~Miquel}
\affiliation{Instituci\'o Catalana de Recerca i Estudis Avan\c{c}ats, E-08010 Barcelona, Spain}
\affiliation{Institut de F\'{\i}sica d'Altes Energies (IFAE), The Barcelona Institute of Science and Technology, Campus UAB, 08193 Bellaterra (Barcelona) Spain}
\author{R.~L.~C.~Ogando}
\affiliation{Laborat\'orio Interinstitucional de e-Astronomia - LIneA, Rua Gal. Jos\'e Cristino 77, Rio de Janeiro, RJ - 20921-400, Brazil}
\affiliation{Observat\'orio Nacional, Rua Gal. Jos\'e Cristino 77, Rio de Janeiro, RJ - 20921-400, Brazil}
\author{A.~Palmese}
\affiliation{Fermi National Accelerator Laboratory, P. O. Box 500, Batavia, IL 60510, USA}
\author{A.~A.~Plazas}
\affiliation{Department of Astrophysical Sciences, Princeton University, Peyton Hall, Princeton, NJ 08544, USA}
\author{A.~K.~Romer}
\affiliation{Department of Physics and Astronomy, Pevensey Building, University of Sussex, Brighton, BN1 9QH, UK}
\author{A.~Roodman}
\affiliation{Kavli Institute for Particle Astrophysics \& Cosmology, P. O. Box 2450, Stanford University, Stanford, CA 94305, USA}
\affiliation{SLAC National Accelerator Laboratory, Menlo Park, CA 94025, USA}
\author{E.~S.~Rykoff}
\affiliation{Kavli Institute for Particle Astrophysics \& Cosmology, P. O. Box 2450, Stanford University, Stanford, CA 94305, USA}
\affiliation{SLAC National Accelerator Laboratory, Menlo Park, CA 94025, USA}
\author{C.~S{\'a}nchez}
\affiliation{Department of Physics and Astronomy, University of Pennsylvania, Philadelphia, PA 19104, USA}
\author{E.~Sanchez}
\affiliation{Centro de Investigaciones Energ\'eticas, Medioambientales y Tecnol\'ogicas (CIEMAT), Madrid, Spain}
\author{V.~Scarpine}
\affiliation{Fermi National Accelerator Laboratory, P. O. Box 500, Batavia, IL 60510, USA}
\author{S.~Serrano}
\affiliation{Institut d'Estudis Espacials de Catalunya (IEEC), 08034 Barcelona, Spain}
\affiliation{Institute of Space Sciences (ICE, CSIC),  Campus UAB, Carrer de Can Magrans, s/n,  08193 Barcelona, Spain}
\author{I.~Sevilla-Noarbe}
\affiliation{Centro de Investigaciones Energ\'eticas, Medioambientales y Tecnol\'ogicas (CIEMAT), Madrid, Spain}
\author{M.~Smith}
\affiliation{School of Physics and Astronomy, University of Southampton,  Southampton, SO17 1BJ, UK}
\author{M.~Soares-Santos}
\affiliation{Brandeis University, Physics Department, 415 South Street, Waltham MA 02453}
\author{F.~Sobreira}
\affiliation{Instituto de F\'isica Gleb Wataghin, Universidade Estadual de Campinas, 13083-859, Campinas, SP, Brazil}
\affiliation{Laborat\'orio Interinstitucional de e-Astronomia - LIneA, Rua Gal. Jos\'e Cristino 77, Rio de Janeiro, RJ - 20921-400, Brazil}
\author{E.~Suchyta}
\affiliation{Computer Science and Mathematics Division, Oak Ridge National Laboratory, Oak Ridge, TN 37831}
\author{M.~E.~C.~Swanson}
\affiliation{National Center for Supercomputing Applications, 1205 West Clark St., Urbana, IL 61801, USA}
\author{G.~Tarle}
\affiliation{Department of Physics, University of Michigan, Ann Arbor, MI 48109, USA}
\author{D.~Thomas}
\affiliation{Institute of Cosmology and Gravitation, University of Portsmouth, Portsmouth, PO1 3FX, UK}
\author{V.~Vikram}
\affiliation{Argonne National Laboratory, 9700 South Cass Avenue, Lemont, IL 60439, USA}
\author{Y.~Zhang}
\affiliation{Fermi National Accelerator Laboratory, P. O. Box 500, Batavia, IL 60510, USA}


\begin{abstract}
In recent years, many \g-ray sources have been identified, yet the unresolved component hosts valuable information on the faintest emission. In order to extract it, a cross-correlation with gravitational tracers of matter in the Universe has been shown to be a promising tool. We report here the first identification of a cross-correlation signal between \g\ rays and the distribution of mass in the Universe probed by weak gravitational lensing. We use the Dark Energy Survey Y1 weak lensing catalogue and the {\it Fermi} Large Area Telescope 9-year \g-ray data, obtaining a signal-to-noise ratio of 5.3. The signal is mostly localised at small angular scales and high \g-ray energies, with a hint of correlation at extended separation. Blazar emission is likely the origin of the small-scale effect. We investigate implications of the large-scale component in terms of astrophysical sources and particle dark matter emission.
\vspace{-3mm}
\end{abstract}

\preprint{FERMILAB-PUB-19-615-AE}
\preprint{DES-2019-0431}

\maketitle

\section{Introduction}
\label{sec:intro}

Astronomy at \g-ray frequencies represents a promising avenue for both astrophysics and particle physics. On one hand, the most violent phenomena in the Universe produce high-energy photons that travel 
all the way to Earth. 
Thus, they bring us information about the physics of rare events such as supernovae, and the behaviour of matter under extreme conditions, as in pulsars and active galactic nuclei (AGNs). On the other hand, the most elusive form of matter in the cosmos---dark matter, which represents about 25\% of all the Universe's energy content---is believed to consist of an exotic fundamental particle, which may annihilate or decay into standard-model particles and thus produce cosmic messengers including \g-ray photons.
In the weakly interacting massive particle (WIMP) scenario, or for any hypothetical dark matter particle with mass in the GeV range or higher, dark matter particle annihilation/decay almost necessarily result in photons at \g-ray energies. Therefore, \g-ray astronomy represents a promising means to investigate the fundamental nature of dark matter. However,
the faintness of the expected emission makes it very difficult to identify such a signal. 

Since 2008, the Large Area Telescope (LAT) mounted on the \textit{Fermi} satellite have been performing the most detailed observations of the extra-galactic \g-ray sky and resolved 5065 \g-ray sources in the energy range $50\,{\rm MeV}$ to $1\, {\rm TeV}$ \cite{2019arXiv190210045T}. Once the point sources and Galactic emission are removed, the remaining \g-ray photons form the so-called unresolved \g-ray background (UGRB).

A method to discriminate between non-thermal \g-ray emission due to astrophysical sources and possible dark matter annihilation/decay in the UGRB has been proposed in Ref.$\,$\cite{Camera:2012cj}. This method relies on cross-correlations of UGRB maps with maps of other tracers of the underlying structure on cosmological scales, such as the weak gravitational lensing effect, or the clustering of galaxies and galaxy clusters (see also refs. \cite{Ando2014,Fornengo:2014}) and CMB lensing \cite{Fornengo:2014cya,Feng:2016fkl}. These are direct gravitational probes of matter, most of which is thought to be dark matter. The energy, redshift and scale dependence of the aforementioned cross-correlations have the potential to disentangle signatures due to astrophysics from dark matter (see also Ref. \cite{Camera:2014rja}).
More generally, the method can provide valuable information on the redshift distribution and on the clustering properties of the unresolved \g-ray source populations, including blazars, AGNs and star-forming galaxies.

Since cross-correlations of the UGRB with gravitational lensing have been proposed as a probe, several observational attempts have followed \cite{Shirasaki2014,Shirasaki:2016kol,Troster:2016sgf,Shirasaki:2018dkz}, but none so far has detected the signal. Here, we report the first detection of such a cross-correlation. We used 108-month \g-ray data from \Fermi\ and first year (Y1) shear measurements from the Dark Energy Survey (DES). In the following, we describe details of the analysis and discuss the results.

\begin{figure*}[t]
  \centering
  \includegraphics[width=0.80\textwidth]{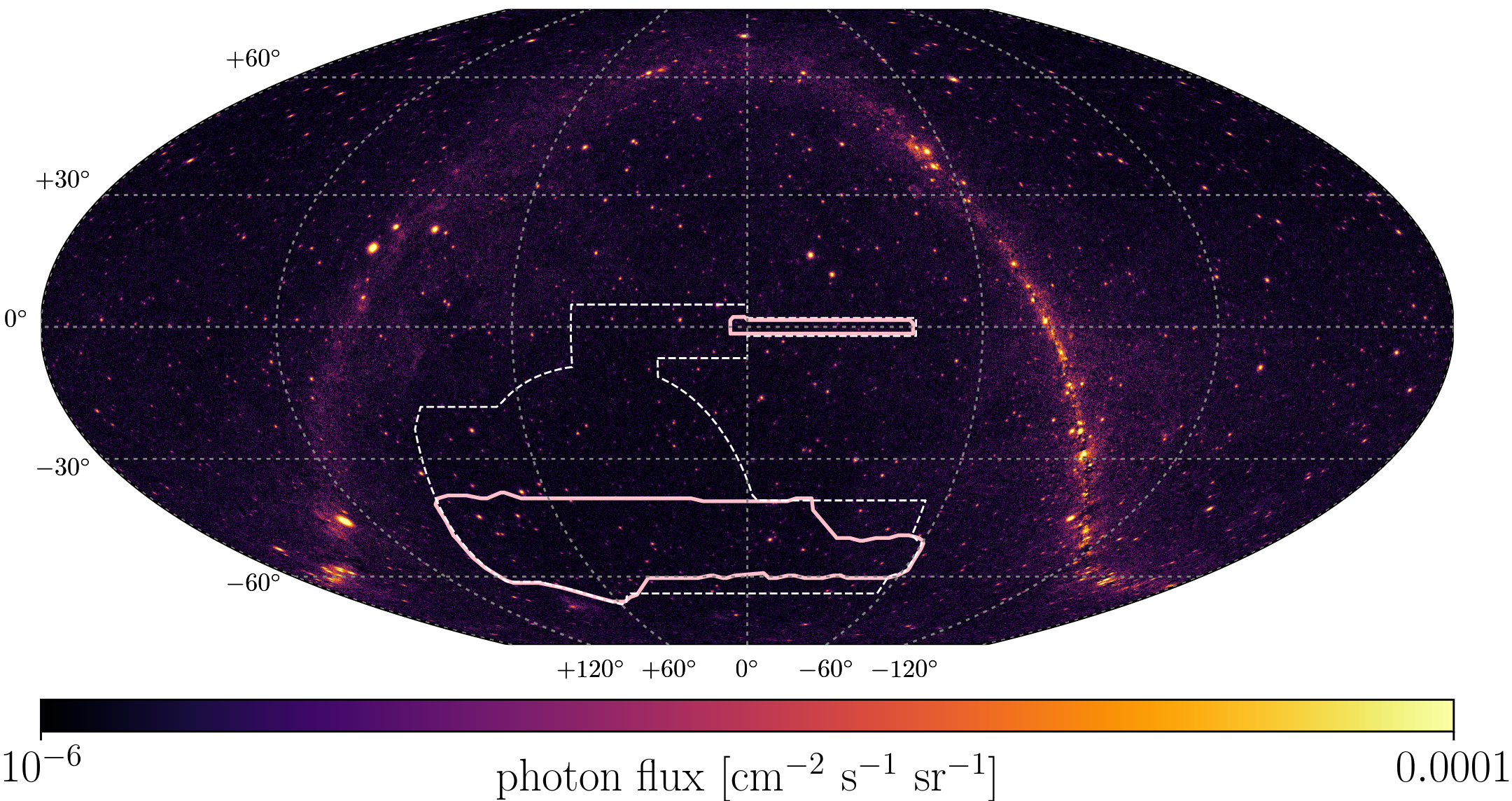}
  \caption{DES Y1 (solid, used in this work) and final (dashed) sky coverage superimposed on the \Fermi\ $\gamma$-ray map for photons in the 1-10 GeV energy range. The Galactic plane and point-source emissions are clearly visible. The plot is in McBryde-Thomas flat polar quartic projection.}
  \label{fig:maps}
\end{figure*}

\section{Analysis and results} 
\label{sec:res}
The observable we probe is the cross-correlation between the unresolved component of the \g-ray emission and gravitational shear. To this aim, the \Fermi\ data have been pre-processed to produce the relevant energy-dependent response functions of the detector and full-sky maps of photon intensities in several energy bins. Resolved \g-ray sources and the bright Galactic plane emission have been masked with energy- and flux-dependent masks, in order to minimise the sky fraction removal. Furthermore, we have subtracted a model of the Galactic plane emission. Galactic foreground emission does not lead to false detection of a cross-correlation, since it does not correlate with the large-scale structure measured by gravitational shear, but it increases the variance of the measurements (see Supplemental Material and, e.g., Refs. \cite{Xia:2015wka,Shirasaki2014,Shirasaki:2016kol,Ammazzalorso:2018evf,Shirasaki:2018dkz}).
The weak lensing information is extracted by measuring the mean tangential ellipticity of source galaxies in the DES footprint around pixels weighted by their UGRB flux.
The shear catalogue is divided in redshift bins in order to perform a tomographic analysis. 
As an illustration of the overlapping area bewteen DES and \Fermi, Fig.~\ref{fig:maps} shows the DES footprint and the \Fermi\ map for photon energies in the 1-10 GeV interval.

We measure the cross-correlation between the UGRB and gravitational shear through its 2-point angular correlation function. Specifically, we adopt the following estimator (see also Ref. \cite{Gruen:2017xjj}):
\begin{align}
\Xi^{ar} (\theta) = \Xi^{\mathrm{signal}}_{\Delta \theta_h, \Delta E_a, \Delta z_r}-\Xi^{\mathrm{random}}_{\Delta \theta_h, \Delta E_a, \Delta z_r} = 
\frac{\sum_{i,j} \, e^r_{ij,\mathrm{t}} \, I^a_j}{R \sum_{i,j} \, I^a_j }-\frac{\sum_{i,j}  \, e^r_{ij,\mathrm{t}} \, I^a_{j, \rm random}}{R \sum_{i,j} I^a_{j, \rm random} } ,
\label{eq:crossshear}
\end{align}
where $ \Xi^{\mathrm{signal}}_{\Delta \theta_h, \Delta E_a, \Delta z_r}$ is the correlation function in configuration space of the two observables measured in different angular ($\Delta \theta_h$), \g-ray energy ($\Delta E_a$) and lensing source-galaxy redshift ($\Delta z_r$) bins. The correlation is obtained by summing the products of tangential ellipticity of source galaxies $i$ relative to a pixel $j$, $e^r_{ij,\mathrm{t}}$, multiplied by the {\it Fermi}-LAT photon intensity flux in the $a$-th energy bin and in pixel $j$, $I^a_j$. The sum runs over all unmasked pixels $j$ and all sources $i$ in the redshift bin of the shear catalogue, and it is performed in each of the different photon energy bins (labelled by $a$) and source galaxies redshift bins (labelled by $r$). Lastly, $R$ is the mean response of ellipticity to shear for sources in the redshift bin, determined by the \textsc{metacalibration} algorithm \cite{Sheldon:2017szh,Zuntz:2017pso} to be between 0.54 and 0.73 for the source-galaxy redshift bins used here.

From the correlation function, we remove $ \Xi^{\mathrm{random}}_{\Delta \theta_h, \Delta E_a, \Delta z_r}$, the measurement of tangential shear around random lines of sight. This is done by setting $I^a_{j, \rm random}=1$ anywhere within the sky region used for \g-ray measurements in that energy bin and 0 elsewhere. This reduces additive shear systematic effects, random very-large-scale structures, or chance shear alignments relative to the mask. The random subtraction, while not affecting the expected signal, lowers the variance at large angular separations (see also Refs. \cite{2017MNRAS.471.3827S,Gruen:2017xjj}). 

\begin{table*}
\centering
\begin{tabular}{cccccccccc}
 \hline
 & \multicolumn{9}{c}{Bin number} \\
 \cline{2-10}
 & 1 & 2 & 3 & 4 & 5 & 6 & 7 & 8 & 9 \\
 \hline
 \hline
$E_{\rm min}$ [GeV] & $0.631$ & $1.202$ & $2.290$ & $4.786$ & $9.120$ & $17.38$ & $36.31$ & $69.18$ & $131.8$ \\
\hline
$E_{\rm max}$ [GeV] & $1.202$ & $2.290$ & $4.786$ & $9.120$ & $17.38$ & $36.31$ & $69.18$ & $131.8$ & $1000.0$ \\
\hline
$\theta_{\rm cont}$ 68\% [deg] & $0.50$ & $0.58$ & $0.36$ & $0.22$ & $0.15$ & $0.12$ & $0.11$ & $0.10$ & $0.10$ \\
 \hline
$\theta_{\rm cont}$ 95\% [deg] & $1.03$ & $1.06$ & $0.62$ & $0.39$ & $0.28$ & $0.22$ & $0.20$ & $0.18$ & $0.16$ \\
 \hline
Photon counts & $345230$ & $444559$ & $286209$ & $102821$ & $41148$ & $16932$ & $5250$ & $1728$ & $722$ \\
 \hline
\end{tabular}
\caption{Energy bins over which the analysis is performed, 68\% and 95\% containment angles $\theta_{\rm cont}$ of the {\it Fermi}-LAT PSF, and photon counts in the unmasked Fermi area in each energy bin}.
\label{tab:enbins}
\end{table*}

We analyse the data in 12 logarithmically-spaced angular bins with radii between $5$ and $600\,\mathrm{arcmin}$, 9 photon energy bins between $0.631$ and $10^3\,\mathrm{GeV}$, and 4 redshift bins defined by
$0.20 < \langle z \rangle < 0.43$, $0.43 < \langle z \rangle < 0.63$, $0.63 < \langle z \rangle < 0.90$ and $0.90 < \langle z \rangle < 1.30$, where $\langle z \rangle$ is the estimated expectation value of galaxy redshift from DES. The energy bins used in the analysis and the corresponding 68\% and 95\% containment angles of the {\it Fermi}-LAT PSF are shown in Tab. \ref{tab:enbins}. These sum up to a total of 432 bins for the cross-correlation measurement. The analysis is performed blindly, i.e.\ on multiple variants of the measurements including artificial versions, in order to avoid experimental bias in measurement and interpretation of the signal. See the Supplemental Material for details.
\begin{figure*}[t]
    \centering
    \includegraphics[width=0.49\textwidth]{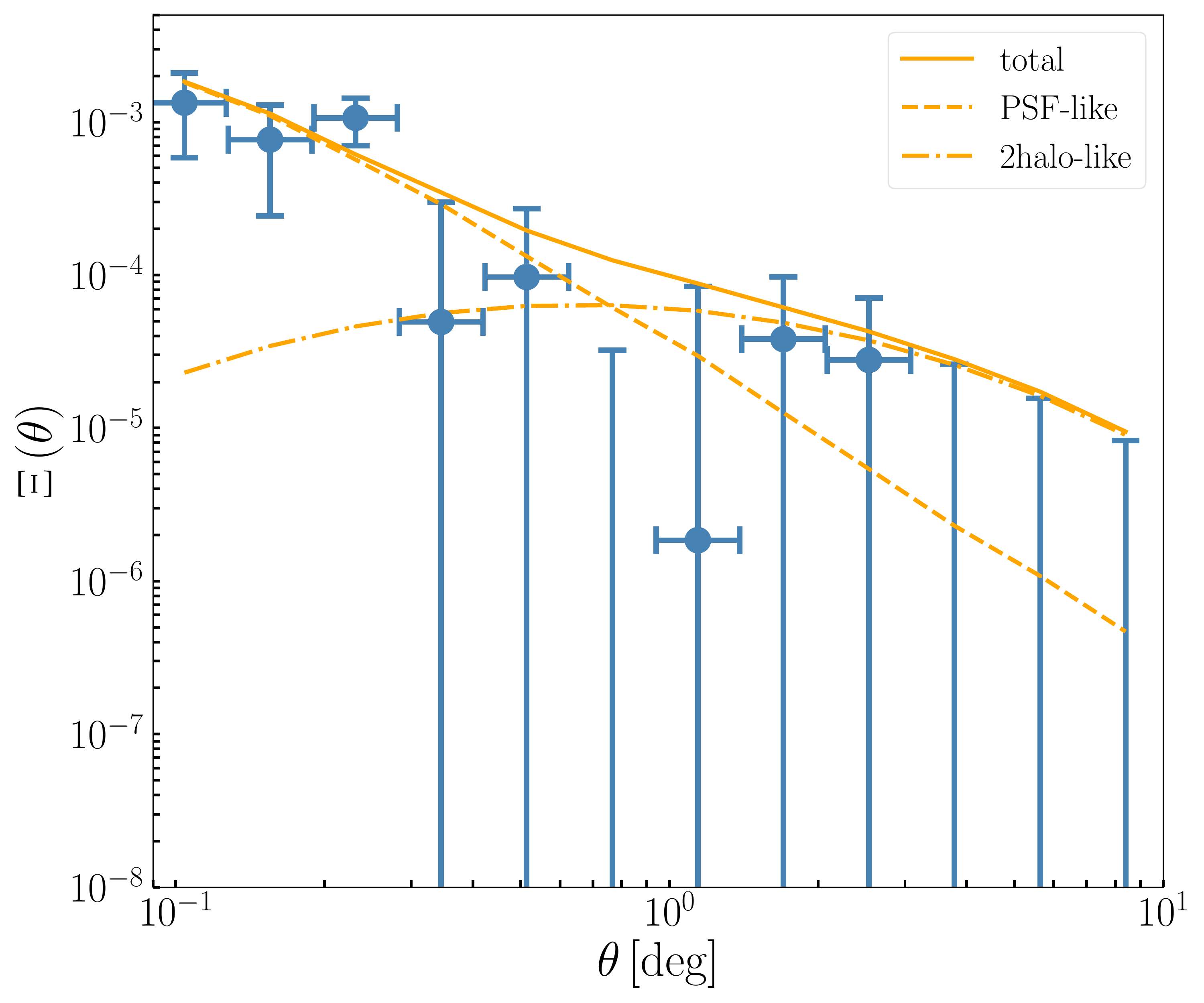}
    \includegraphics[width=0.49\textwidth]{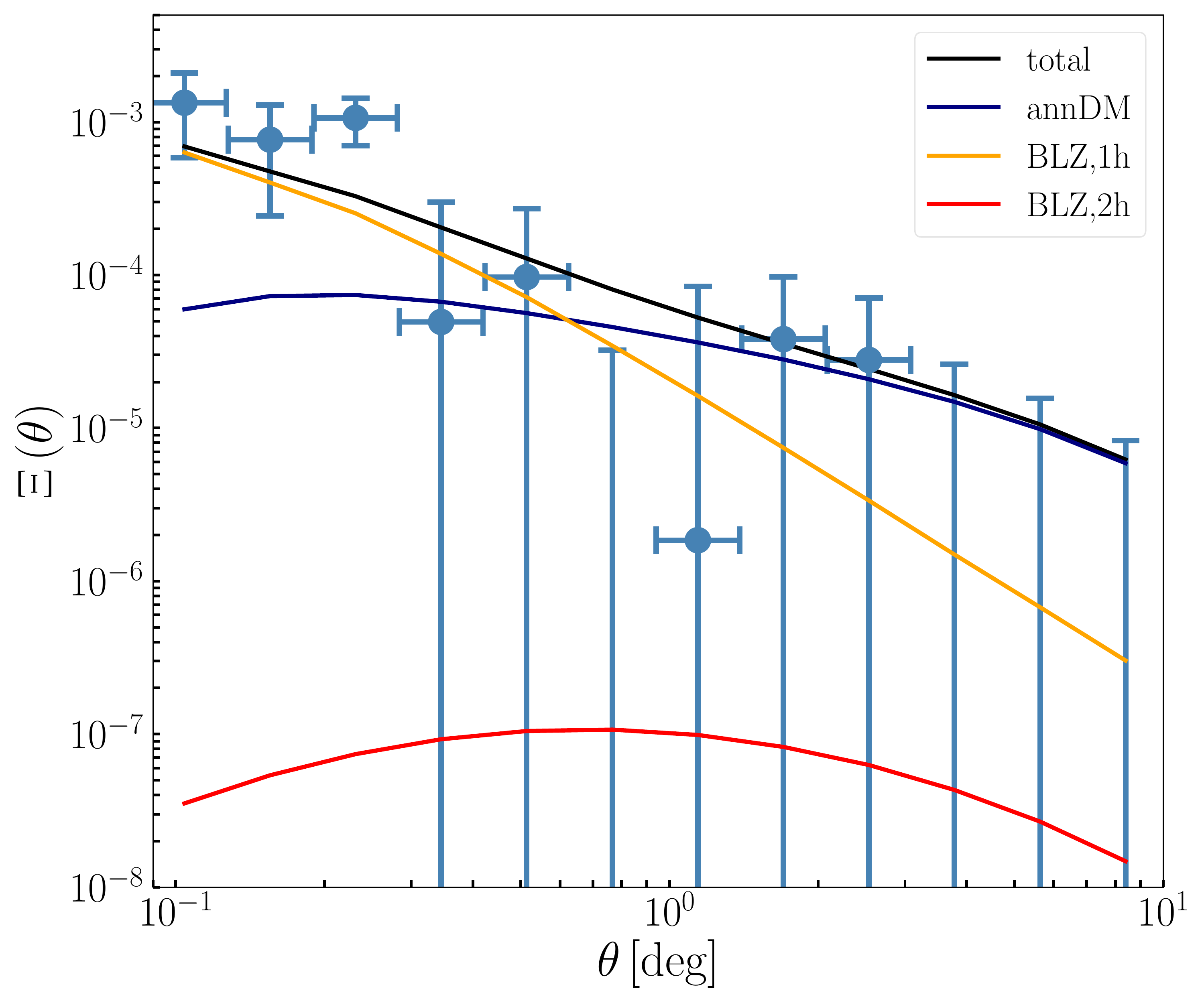}
    \caption{Measurement and model of the cross-correlation between \g-ray photons and gravitational shear. The points in both panels show the measured cross-correlation, averaged over all redshift and energy bins, while the fits is done across all dimensions. The lines refer the best fit results for the \pheno\ model (left) and for the \phys\ model (right), averaged the same way.}
    \label{fig:integratedsignal}
\end{figure*}

The result of the measured cross-correlations, averaged over all energy and redshift bins, is shown in Fig.~\ref{fig:integratedsignal} in terms of the estimator $\Xi(\theta)$ defined in Eq.~(\ref{eq:crossshear}). Note that the data points reported on both panels are the same, although confronted with different models. A clear positive cross-correlation is observed, especially at small angular separations.

In order to determine the statistical significance of the signal, we test the deviation of the measurement from a null signal (null hypothesis of pure noise) by means of a \pheno\ model, which aims at
capturing the general expected features of the cross-correlation signal without resorting to any specific, detailed modelling of its physical origin (in the next Section, we will instead adopt a physical model to provide insights on the origin of the cross-correlation). In the halo-model approach, all mass in the large-scale structure of the Universe is associated with virialised dark matter halos, and the correlation function can thus be decomposed into the so-called 1-halo and a 2-halo terms (`1h' and `2h', in formulae hereafter). The former refers to the correlation between two points in the same physical halo; the latter to the case in which the two points belong to two different halos. Point-like sources contribute at small angular scales with a 1-halo term, while at large scales they produce a 2-halo term resembling the large-scale structure matter distribution. In our case, we use the fact that the spatial extent of the 1-halo term is smaller than the beam window function of the \Fermi. Then a \pheno\ model can be constructed as:
\begin{align}
\Xi_{\rm phe}^{ar}(\theta) \ \langle I_a \rangle = A_1\times E_a^{-\alpha_1}\times (1+z_r)^{\beta_1}\times\hat{\Xi}_\textrm{PSF-like}^{a}(\theta) \, + A_2\times E_a^{-\alpha_2+2.2}\times (1+z_r)^{\beta_2}\times \hat\Xi_\textrm{2h-like}^{ar}(\theta)\,
\label{eq:phenomdl}
\end{align}
where $E_a$ and $z_r$ are the central values of the energy (measured in  GeV) and redshift bins, and $\langle I_a \rangle$ is the measured photon flux. $\hat{\Xi}_\textrm{PSF-like}^{a}(\theta)$ is the Legendre transform of the beam window function (or point-spread function, PSF) integrated in the $a$th energy bin (in arbitrary units, being merely a template for the 1-halo term due to point-like $\gamma$-ray sources) and $\hat\Xi_\textrm{2h-like}^{ar}(\theta)$ is the Legendre transform of a generic 2-halo (i.e.\ large-scale) contribution, also convolved with the \Fermi\ beam window function. 
Correlation functions with a hat have flux units, while those without a hat are normalised to the \g-ray flux as in Eq. (\ref{eq:crossshear}), and therefore dimensionless. The two normalisations $A_1$ and $A_2$, spectral indices $\alpha_1$ and $\alpha_2$, and redshift evolution indices $\beta_1$ and $\beta_2$ are free parameters of the model \footnote{For blazars, we expect an average spectral index of $2.2$, which is the reason for the term added to $\alpha_2$.}.
Gamma-ray sources typically have energy spectra that can be well approximated by a power-law, and so it is assumed in Eq. (\ref{eq:phenomdl}). For simplicity, we also assume a power-law scaling in redshift. Best fits and confidence intervals of the parameters are found in a  Markov Chain Monte Carlo likelihood analysis.

The first statistical method adopted to quantify the presence of a signal, and its significance, against the null hypothesis relies on the $\Delta\chi^2$ test statistics, with the chi-squared defined in the usual way, i.e.\ : 
%
%
\begin{align}
\chi^2({\bm P}_{\rm mod}) = [{\bm \Xi}_{\rm data} -{\bm \Xi}_{\rm th}({\bm P}_{\rm mod})]^{\sf T}{\bm \Gamma}^{-1}[{\bm \Xi}_{\rm data} -{\bm \Xi}_{\rm th}({\bm P}_{\rm mod})]
\end{align}
where ${\bm \Xi}_{\rm data}$ is the data vector, ${\bm \Xi}_{\rm th}$ is the theoretical cross-correlation for the models outlined above, described by the parameter set ${\bm P}_{\rm mod}$, and ${\bm \Gamma}$ is the data covariance matrix, detailed in the Appendix.
(All angular, energy, and redshift bin indexes have been omitted for simplicity of notation.) The $\Delta\chi^2$ is defined as $\Delta\chi^2_{\rm mod}=\chi^2_{\rm null}- \chi^2({\bm P}_{\rm mod}^\star)$, with $\chi^2({\bm P}_{\rm mod}^\star)$ computed at the model parameter values ${\bm P}_{\rm mod}^\star$ that best fit the data, and $\chi^2_{\rm null}$ referring to no signal, i.e.\ ${\bm \Xi}_{\rm th}=0$. The second estimator is the matched filter signal-to-noise ratio (see e.g. Ref. \cite{Becker:2015ilr}),
\be
{\rm SNR}({\bm P}_{\rm mod})=\frac{{\bm \Xi}_{\rm data}^{\sf T}{\bm \Gamma}^{-1}{\bm \Xi}_{\rm th}({\bm P}_{\rm mod})}{\sqrt{{\bm \Xi}_{\rm th}^{\sf T}({\bm P}_{\rm mod}){\bm \Gamma}^{-1} {\bm \Xi}_{\rm th}({\bm P}_{\rm mod})}};
\ee
in analogy to $\Delta\chi^2_{\rm mod}$, we shall later refer to ${\rm SNR}_{\rm mod}\equiv{\rm SNR}({\bm P}_{\rm mod}^\star)$.

In Table~\ref{tab:chi2comp} we present the results on detection significance. The \pheno\ model results for the full data set show a clear evidence for the presence of a cross-correlation signal, at the level of $\mathrm{SNR}_{\rm mod}=5.3$.  Since the matched filter based on the \pheno\ model captures the generic features of the cross-correlation signal, without committing to any specific physical description, this best assesses that indeed a cross-correlation between gravitational shear and unresolved \g-rays emission has been observed. In order to investigate the features of the signal in more detail, we repeat the tests by subdividing the data set according to redshift, energy, and angular separation. Specifically, Low/High-$z$ refers to the first two and second two redshift bins; Low/High-$E$ bins are defined by being below/above $5\,\mathrm{GeV}$; and Small/Large-$\theta$ separates angular scales below/above 3 times the \Fermi\ PSF.
\begin{table}
\centering
\begin{tabular}{cccccccc}
\hline
 & \multicolumn{7}{c}{Data set} \\
 \cline{2-8}
 & Full & Low-$z$ & High-$z$ & Low-$E$ & High-$E$ & Small-$\theta$ & Large-$\theta$ \\
\hline
\hline
$\Delta \chi^2_{\rm phe}$ & $27.5$ & $17.5$ & $10.4$ & $1.7$ & $21.7$ & $17.0$ & $6.0$ \\
${\rm SNR}_{\rm phe}$ & $5.3$ & $4.2$ & $3.2$ & $1.3$ & $5.1$ & $4.2$ & $2.5$ \\
\hline
$\Delta \chi^2_{\rm phys}$ & $27.0$ & $4.8$ & $12.5$ & $1.5$ & $16.2$ & $12.5$ & $4.8$ \\
${\rm SNR}_{\rm phys}$ & $5.2$ & $2.2$ & $3.5$ & $1.2$ & $4.2$ & $3.5$ & $2.1$ \\
 \hline
\end{tabular}
\caption{$\Delta \chi^2_{\rm mod}$ and ${\rm SNR}_{\rm mod}$ computed for the \pheno\ and \phys\ models, using either the full data set or the various subsamples discussed in the text. For dark matter in the \phys\ model, we consider the annihilation channel $\tau^+\tau^-$. For the Low-$z$ case we selected the two first redshift bins, while for the High-$z$ case the last two bins, where the bins are defined as: $0.20 < \langle z \rangle < 0.43$, $0.43 < \langle z \rangle < 0.63$, $0.63 < \langle z \rangle < 0.90$ and $0.90 < \langle z \rangle < 1.30$; the Low-$E$ subset is defined for energies below 5 GeV, while the High-$E$ for energies above this value; finally, the {Small-$\theta$/Large-$\theta$} cases correspond to data points below/above 3 times the \Fermi\ PSF.}
\label{tab:chi2comp}
\end{table}

From Table~\ref{tab:chi2comp} we infer that the signal is mostly concentrated at high energies and small angles.
These results point towards an interpretation in terms of point-like sources with hard energy spectrum, broadly compatible with these sources being blazars.
In fact, the best-fit for the spectral index of the PSF-like 1-halo component $\alpha_1=1.81^{+0.20}_{-0.24}$ is quite hard with respect to the spectral index of the average intensity of the UGRB \cite{Ackermann:2014usa}, but compatible with BL Lac emission, which is the source population expected to be the most relevant in the range of fluxes probed by this analysis, just below the \Fermi\ flux sensitivity threshold.  Notice that this hard spectral index is in agreement with recent findings from \g-ray auto-correlation analysis \cite{Ackermann:2018wlo}, possibly suggesting that BL Lac objects below the threshold have slightly harder spectra than those detected individually.
The energy scaling of the 2-halo component is also compatible with a blazar origin, though this term shows lower statistical significance than the 1-halo component.
Concerning the redshift dependence of the signal, the statistical significance is almost equally distributed among the lower and higher redshift bins. 
The allowed regions for the parameters of the {\it phenomenological } model are shown in Fig. \ref{fig:triangular}, while the cross-correlation function for the best-fit of the \pheno\ model are shown in the left panel of Fig.~\ref{fig:integratedsignal}: the PSF-like term due to point-like sources well reproduces the behaviour of the measured cross-correlation up to about $1\,\deg$ scale.
We note here that for the subset of High-$E$/Small-$\theta$, comprising 88 data-points, we do obtain a distinctive signal without application of the matched filter. The $\chi^2_{\rm null}=137$ for these points corresponds to a p-value of 0.0006, meaning that the null hypothesis is excluded at $3.5\sigma$ in this subset.

\section{Discussion} 
\label{sec:disc}

\label{sec:phys}
{In the following we attempt a physical interpretation of the signal detected in the previous Section.}
Star-forming galaxies and misaligned AGNs are not expected to be able to produce a sufficiently hard energy spectrum, which thus points to a dominant blazar component. Particle dark matter in terms of WIMPs can also provide a hard spectrum, especially if the annihilation channel is predominantly leptonic or, in the case of a hadronic final state, if the dark matter mass is large (above a few hundred GeV).

Blazars are compact sources and, for our purposes, they can be considered as point-like---i.e.\ their size is, on average, much smaller than the \Fermi\ PSF. Also the size of the halo hosting blazars rarely exceeds the \Fermi\ PSF. This has a consequence that the angular correlation function for the 1-halo term essentially follows from the detector PSF. Conversely, the relevant dark matter halos have a larger angular extent, and the corresponding 1-halo correlation function thus drops more slowly with angular scale. On very large scales, the correlation functions of the two components have  a similar angular behaviour, since the 2-halo power spectra differ only by the bias terms. The fact that our signal is detected with high significance only on small scales therefore points towards blazars as the dominant source. In order to investigate this interpretation, we perform the statistical tests discussed in the previous Section with a \phys\ model, based on a 
detailed characterisation of the components expected to produce the cross-correlation signal: blazars (BLZ), misaligned active galactic nuclei (mAGN), star-forming galaxies (SFG) and possibly dark matter (DM). 
The {\it physical} cross-correlation function model reads:
\begin{align}
\Xi_{\rm phys}^{ar}(\theta) \ \langle I_a \rangle =  A_{\rm BLZ}^{\rm 1h} \times \hat\Xi_{\rm BLZ,1h}^{ar} (\theta)+  A_{\rm BLZ}^{\rm 2h} \times \hat\Xi_{\rm BLZ,2h}^{ar} (\theta) +  
A_{\rm mAGN} \times \hat\Xi_{\rm mAGN}^{ar} (\theta)+ A_{\rm SFG} \times \hat\Xi_{\rm SFG}^{ar} (\theta) + 
A_{\rm DM} \times \hat\Xi_{ {\rm DM}}^{ar}(\theta; \mdm) \, .
\label{eq:physmdl}
\end{align}
The model parameters are: free normalisations for the astrophysical sources, $A_{\rm BLZ}^{\rm 1h}$, $A_{\rm BLZ}^{\rm 2h}$, $A_{\rm mAGN}$, and $A_{\rm SFG}$; the mass of the dark matter particle, $m_{\rm DM}$; its velocity-averaged annihilation rate, $\sv$,  expressed in terms of the ``thermal'' cross-section $\sv_{\rm th} = 3 \times 10^{-26}\, {\rm cm^{3}\, s^{-1}}$ through the normalisation $A_{\rm DM} \equiv \sv / \sv_{\rm th}$. 
Note that for blazars, which represent the astrophysical component expected to dominate the correlation signal at the current level of unresolved \g-ray emission, we allow the 1-halo and the 2-halo terms to be separately adjusted in the fit against the data. 
As for the \pheno\ model, all terms depend on both energy and redshift, labelled by indices $a$ and $r$, respectively. 

The results are shown in Table~\ref{tab:chi2comp}, where the overall significance of the presence of a signal, the preference for an origin at high energies and small angular scales are all confirmed. However, since in this case we have specific behaviours for the correlation functions as dictated by a physical model (different for each component, contrarily to the average generic case of the \pheno\ model), we notice that a mild hint of large scale correlation is present---namely, in the Large-$\theta$ case. We note that both the physical and phenomenological models provide a good fit to the data according to their $\chi^2$ (see the Supplemental Material).

\begin{figure*}[t]
    \centering
    \includegraphics[width=0.45\textwidth]{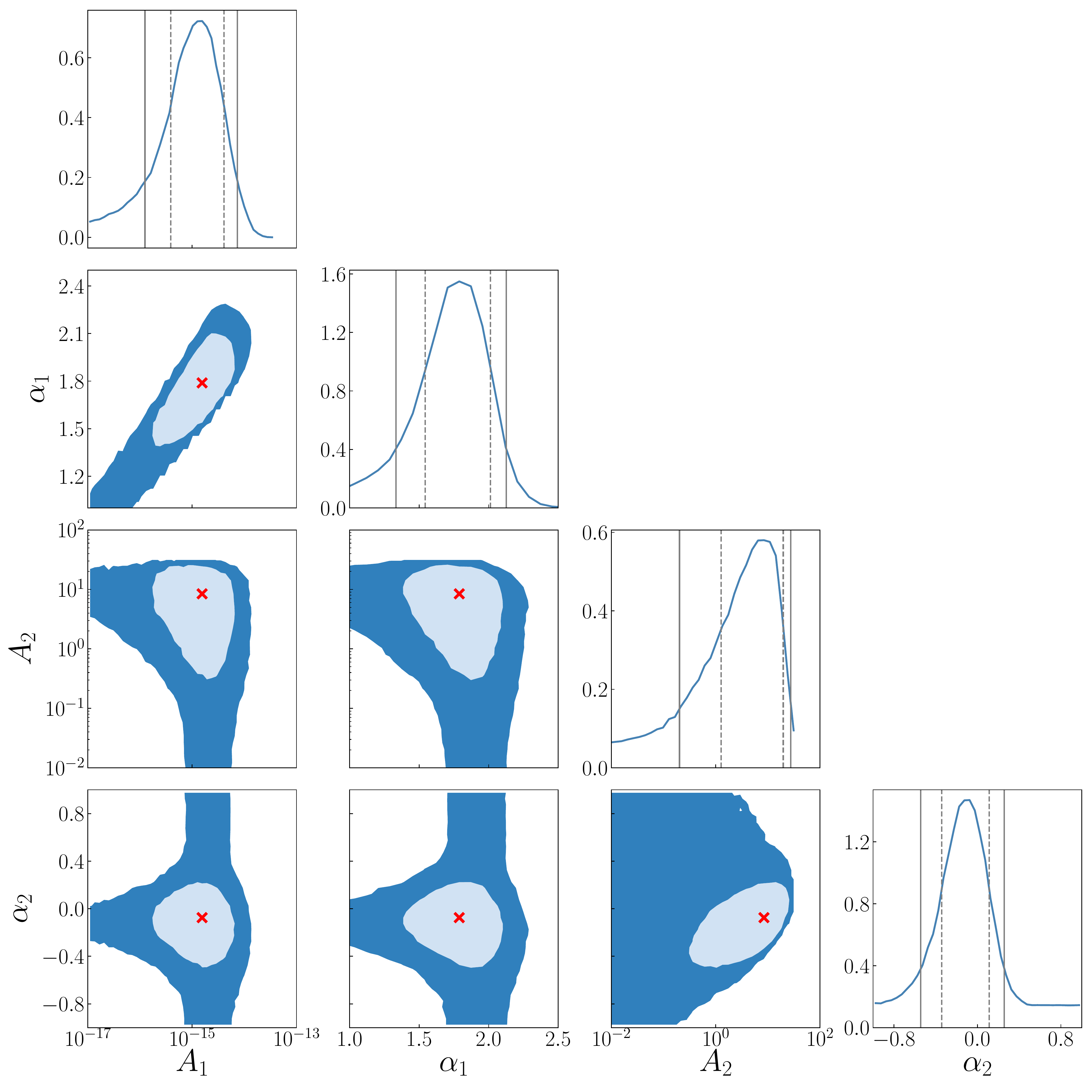}
    \includegraphics[width=0.45\textwidth]{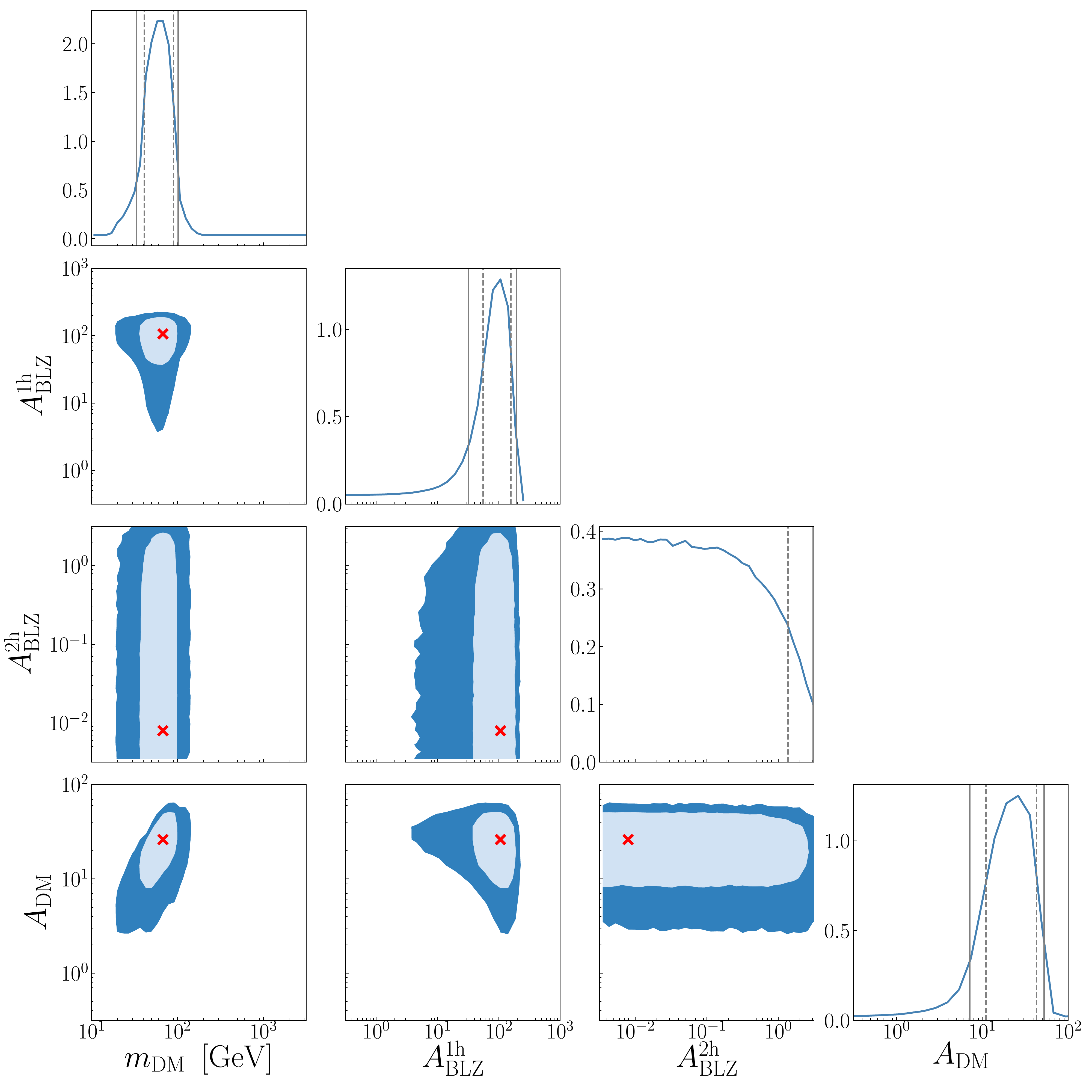}
   \caption{Left: Constraints on the normalisation and spectral index parameters of the {\it phenomenological} model (the redshift dependence parameters are unconstrained and not shown in the plot). Right: Constraints on the parameters of the dark matter and blazar models described in Eq.~\ref{eq:physmdl}. The blazar model assumes a single population matching the properties of {\it Fermi} resolved sources. The dark matter model assumes annihilation in the $\tau^+\tau^-$ channel. In both panels, 2D contours
    refer to the $68\%$ and $95\%$ C.L. regions.
        The dashed and solid vertical lines in the 1D subplots denote the the $68\%$ and $95\%$ C.L. constraints of the 1D profile likelihood distributions.}
    \label{fig:triangular}
\end{figure*}

More details of the analysis are shown in Fig.~\ref{fig:triangular}, where the triangular plot of the profile likelihood distributions of the model parameters is reported. The likelihood exhibits a preference for a large 1-halo term of blazars with normalisation $A_{\rm BLZ}^{\rm 1h}=102^{+56}_{-57}$, while the normalisations of the blazar 2-halo term shows only a (weak) upper bound. The latter picture is shared also by the other astrophysical sources (SFG and mAGN) which are shown only in the Appendix for brevity. 

The blazar-shear cross-correlation on small scales depends on the relation between the blazar \g-ray luminosity and the host-halo mass, a quantity which is rather uncertain. For our reference model this relation has been taken from \cite{Camera:2014rja}, where it was derived by associating the \g-ray luminosity of blazars to the mass of the supermassive black hole powering the AGN and then relating the mass of the black hole to the mass of the dark matter halo. This procedure gives $M(\mathcal{L})=2\times 10^{13}M_\odot\left[\mathcal{L}/(10^{47}\,\mathrm{erg\,s^{-1}})\right]^{0.23}(1+z)^{-0.9}$, where $\mathcal{L}$ is the rest-frame luminosity of blazars in the energy range $0.1$ to $100\,\mathrm{GeV}$. We can therefore translate a value of $A_{\rm BLZ}^{\rm 1h}$ different from unity to a deviation from the reference $M(\mathcal{L})$ relation. The value we found implies that the average mass of a halo hosting an unresolved blazar is larger than the one adopted in Ref. \cite{Camera:2014rja}, and most likely above $10^{14}M_\odot$. The cross-correlation signal with weak lensing seems therefore to be dominated by blazars residing in cluster-size halos.

The right panel of Fig.~\ref{fig:integratedsignal} shows that the cross-correlation at small angular scales requires a sizeable blazar 1-halo term. It also illustrates that the best-fit analysis exhibits a mild preference for some power at large scales. This can be accounted for either by the 2-halo term of blazars or by a DM contribution. The interplay of the angular, energy and redshift behaviours of the observed signal leads to a small preference for a DM component over a pure blazar contribution in the assumed model. Misaligned AGNs and star-forming galaxies are disfavoured since they do not meet the requirement of a hard energy spectrum. 

\begin{figure*}[t]
    \centering
    \includegraphics[width=0.45\textwidth]{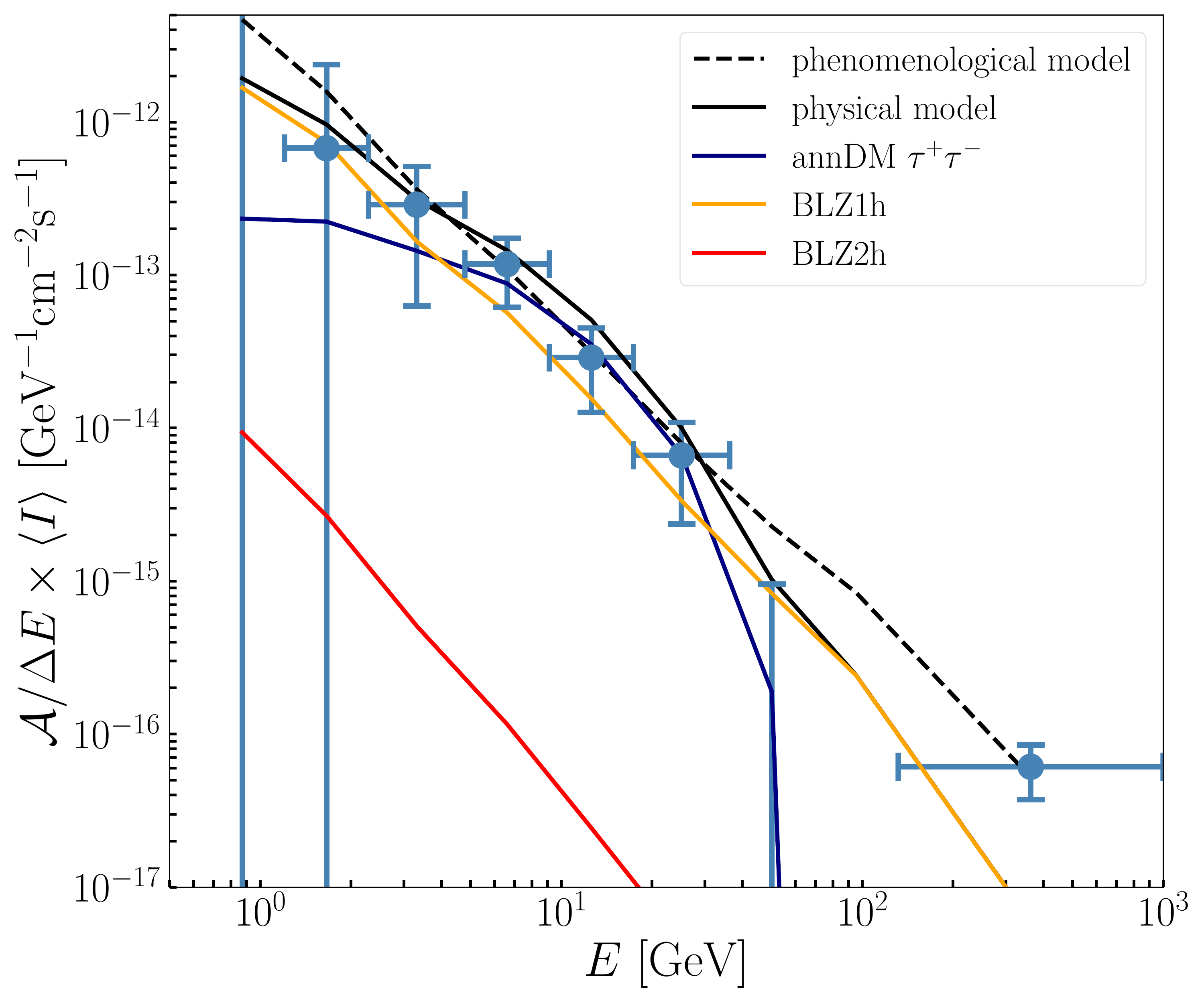}
    \includegraphics[width=0.45\textwidth]{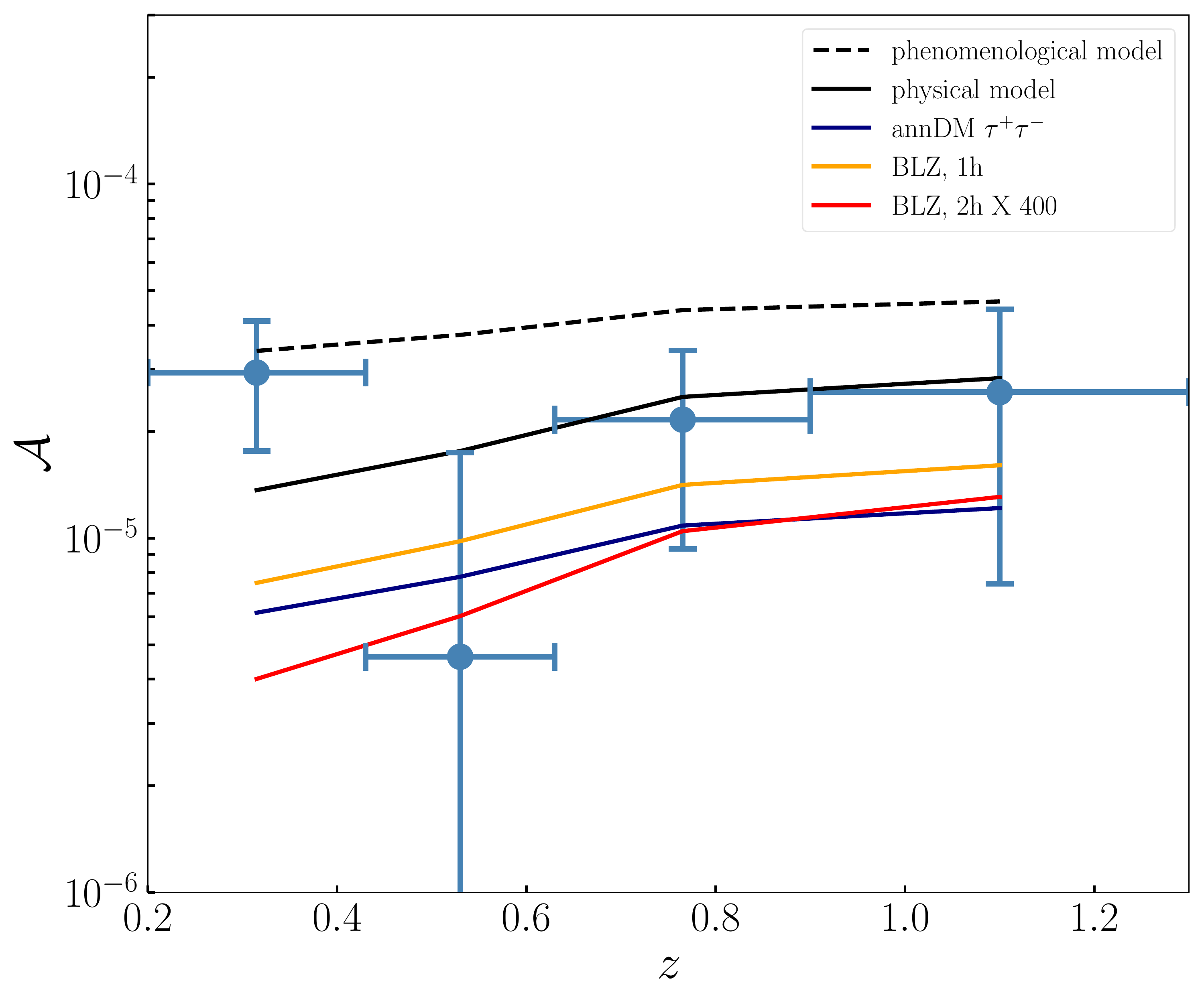}
    \caption{Left: Energy scaling of the measured signal and best-fit models, in terms of the matched filter amplitude ${\cal A}$ of the cross-correlation between gravitational shear and \g\ rays, see text for its definition. The amplitude is divided by the size of the corresponding energy bin $\Delta E_a$ and multiplied by the measured photon intensity $\langle I_a \rangle$ in the same bin, to show the physical differential scaling in energy. Right: Redshift scaling of the measured signal and models, again in terms of the matched filter amplitude ${\cal A}$ introduced in the text.}
    \label{fig:energyredshift}
\end{figure*}

To have a visual impression on the physical behaviours, we plot the energy and redshift dependence of the cross-correlation signal in Fig.~\ref{fig:energyredshift}. The average along the angular and redshift (energy) directions of each point of the energy (redshift) spectrum is performed by computing a matched filter amplitude ${\cal A}={\bm \Xi}^{\sf T}{\bm \Gamma}^{-1}\bar {\bm \Xi}_{\rm M}/(\bar {\bm \Xi}_{\rm M}^{\sf T}{\bm \Gamma}^{-1} \bar {\bm \Xi}_{\rm M})$, where $\bar {\bm \Xi}_{\rm M}$ is given by a sample model that we choose to be flat in energy and redshift, while scaling as $1/\theta$ in angle, to approximately reproduce the expected signal, and ${\bm \Xi}$ is either the measured data or the best-fit models introduced in the main text. The error on ${\cal A}$ is given by $\sigma^2_{\cal A}=({\bm \Gamma}^{-1} \bar {\bm \Xi}_{\rm M})^{\sf T}{\bm \Gamma}({\bm \Gamma}^{-1}\bar {\bm \Xi}_{\rm M})/(\bar {\bm \Xi}_{\rm M}^{\sf T}{\bm \Gamma}^{-1} \bar {\bm \Xi}_{\rm M})^2$.
From Fig.~\ref{fig:energyredshift} one can appreciate the small but noticeable differences in the energy and redshift scalings of the models of different physical components that have been just discussed.

Notice that the blazar model we are adopting, and which is outlined in the Supplemental Material, is based on the current understanding of blazars as derived from the {\it Fermi}-LAT resolved sources: the small preference in the fit for a contribution with features compatible with DM might be interpreted as an indication that unresolved blazars have different properties than the resolved ones.

We find that
for a dark matter particle dominantly annihilating into the leptonic channel $\tau^+\tau^-$ the best-fit improves by $2.8\sigma$ as compared to a model where DM is not included. The lack of a degeneracy between dark matter and blazar amplitudes (see Fig.~\ref{fig:triangular}) indicates that the two components are supported by independent features of our cross-correlation data, in particular the small and large scale behaviour. The best-fit occurs for a dark matter particle of mass $\mdm = (65 \, \pm \, _{23}^{27})\,\mathrm{GeV}$ and annihilation rate $ \sv = (26 \, \pm \, _{15}^{17}) \times \svth $. In the case of a softer energy spectrum, as the one provided by a $\bar b b$  annihilation channel, the fit improvement is slightly lower, at $2.7\sigma$ level, with best-fit mass $\mdm = (302 \, \pm \, _{120}^{188})\,\mathrm{GeV}$ and annihilation rate $ \sv = (78 \, \pm \, _{43}^{67}) \times \svth $.


Let us remark that the main source of uncertainty concerning the dark matter signal described in this work is our ignorance on the impact of substructures. Specifically, the minimal halo mass and the amount and distribution of sub-haloes can significantly change the size of the signal. This, however, is common to all cosmological searches for a particle dark matter signature. Comparing our nominal model \cite{Moline2016} to a recent development \cite{Hiroshima:2018kfv} based on both $N$-body simulations and analytical modelling, we found that the constraint on the annihilation cross-section gets shifted to higher annihilation rates by about one order of magnitude. This is due to a different amplitude of the expected signal, whereas the energy, redshift, and angular dependencies are only slightly modified, and the predicted cross-correlation function therefore just needs a larger normalisation, which in turn is directly provided by the annihilation rate.

We conclude by summarizing the major results of this analysis. We present the first detection of the cross-correlation between the \g-ray sky and the mass distribution in the Universe observed through gravitational lensing shear, with a significance of SNR$=5.3$. The bulk of this signal comes from the 1-halo term of point-like sources with a hard spectrum, most likely dominated by blazars. In addition, we find a hint 
for a cross-correlation on large scales with spectral and redshift behaviours that might imply that the population of blazars that are currently unresolved by {\it Fermi}-LAT has different characteristics than those obtained by extrapolation from observations of resolved blazars, or that an additional contributor to the \g-ray emission is present.
The analysis of the cross-correlation of \Fermi\ data with the forthcoming Year 3 and Year 5 DES data set, and improvements in modelling of the blazar population will likely clarify the source of the signal detected in this work, and characterise it more deeply.

\section{Acknowledgments}

We warmly acknowledge Mattia Fornasa, for his collaboration in the early phases of this project.
This work  is supported by:  `Departments of Excellence 2018-2022' grant awarded by the Italian Ministry of Education, University and Research (\textsc{miur}) L.\ 232/2016; Research grant `The Anisotropic Dark
Universe' No.\ CSTO161409, funded by Compagnia di Sanpaolo and University of Turin; Research grant TAsP (Theoretical Astroparticle Physics) funded \textsc{infn}; Research grant `The Dark Universe: A Synergic Multimessenger Approach' No.\ 2017X7X85K funded by \textsc{miur}; 
Research grant `From Darklight to Dark Matter: understanding the galaxy/matter  connection to measure the Universe' No.\ 20179P3PKJ funded by \textsc{miur}; Research grant `Deciphering the high-energy sky via cross correlation' funded by Accordo Attuativo ASI-INAF n.2017-14-H.0. 
 This work was supported by the Department of Energy, Laboratory Directed Research and Development program at SLAC National Accelerator Laboratory, under contract DE-AC02-76SF00515 and as part of the Panofsky Fellowship awarded to DG. Support for DG was also provided by Chandra Award Number GO8-19101A, issued by the Chandra X-ray Observatory Center. SC is supported by \textsc{miur} through Rita Levi Montalcini project `\textsc{prometheus} -- Probing and Relating Observables with Multi-wavelength Experiments To Help Enlightening the Universe's Structure'. SA was supported by JSPS KAKENHI Grant Numbers JP17H04836, JP18H04578, and JP18H04340. This work was supported in part by the U.S. Department of Energy under 
contract number DE-AC02-76SF00515.

Funding for the DES Projects has been provided by the U.S. Department of Energy, the U.S. National Science Foundation, the Ministry of Science and Education of Spain,
the Science and Technology Facilities Council of the United Kingdom, the Higher Education Funding Council for England, the National Center for Supercomputing
Applications at the University of Illinois at Urbana-Champaign, the Kavli Institute of Cosmological Physics at the University of Chicago,
the Center for Cosmology and Astro-Particle Physics at the Ohio State University,
the Mitchell Institute for Fundamental Physics and Astronomy at Texas A\&M University, Financiadora de Estudos e Projetos,
Funda{\c c}{\~a}o Carlos Chagas Filho de Amparo {\`a} Pesquisa do Estado do Rio de Janeiro, Conselho Nacional de Desenvolvimento Cient{\'i}fico e Tecnol{\'o}gico and
the Minist{\'e}rio da Ci{\^e}ncia, Tecnologia e Inova{\c c}{\~a}o, the Deutsche Forschungsgemeinschaft and the Collaborating Institutions in the Dark Energy Survey.

The Collaborating Institutions are Argonne National Laboratory, the University of California at Santa Cruz, the University of Cambridge, Centro de Investigaciones Energ{\'e}ticas,
Medioambientales y Tecnol{\'o}gicas-Madrid, the University of Chicago, University College London, the DES-Brazil Consortium, the University of Edinburgh,
the Eidgen{\"o}ssische Technische Hochschule (ETH) Z{\"u}rich,
Fermi National Accelerator Laboratory, the University of Illinois at Urbana-Champaign, the Institut de Ci{\`e}ncies de l'Espai (IEEC/CSIC),
the Institut de F{\'i}sica d'Altes Energies, Lawrence Berkeley National Laboratory, the Ludwig-Maximilians Universit{\"a}t M{\"u}nchen and the associated Excellence Cluster Universe,
the University of Michigan, the National Optical Astronomy Observatory, the University of Nottingham, The Ohio State University, the University of Pennsylvania, the University of Portsmouth,
SLAC National Accelerator Laboratory, Stanford University, the University of Sussex, Texas A\&M University, and the OzDES Membership Consortium.

Based in part on observations at Cerro Tololo Inter-American Observatory, National Optical Astronomy Observatory, which is operated by the Association of
Universities for Research in Astronomy (AURA) under a cooperative agreement with the National Science Foundation.

The DES data management system is supported by the National Science Foundation under Grant Numbers AST-1138766 and AST-1536171.
The DES participants from Spanish institutions are partially supported by MINECO under grants AYA2015-71825, ESP2015-66861, FPA2015-68048, SEV-2016-0588, SEV-2016-0597, and MDM-2015-0509,
some of which include ERDF funds from the European Union. IFAE is partially funded by the CERCA program of the Generalitat de Catalunya.
Research leading to these results has received funding from the European Research
Council under the European Union's Seventh Framework Program (FP7/2007-2013) including ERC grant agreements 240672, 291329, and 306478.
We  acknowledge support from the Australian Research Council Centre of Excellence for All-sky Astrophysics (CAASTRO), through project number CE110001020, and the Brazilian Instituto Nacional de Ci\^encia
e Tecnologia (INCT) e-Universe (CNPq grant 465376/2014-2).

This manuscript has been authored by Fermi Research Alliance, LLC under Contract No. DE-AC02-07CH11359 with the U.S. Department of Energy, Office of Science, Office of High Energy Physics. The United States Government retains and the publisher, by accepting the article for publication, acknowledges that the United States Government retains a non-exclusive, paid-up, irrevocable, world-wide license to publish or reproduce the published form of this manuscript, or allow others to do so, for United States Government purposes.

This paper has gone through internal review by the DES collaboration.
%

\appendix

\section{Gravitational lensing data} \label{sec:DESdata}
The Dark Energy Survey (DES \cite{2016MNRAS.460.1270D}) is a photometric survey
performed with the Dark Energy Camera (DECam \cite{2015AJ....150..150F}) on the Blanco
$4\,\mathrm m$ telescope at Cerro Tololo Inter-American Observatory (CTIO) in Chile. DES has observed approximately $5000\,\deg^2$ of the Southern sky in five broadband filters, \textit{g}, \textit{r}, \textit{i}, \textit{z}, and \textit{Y}, ranging from $400$ to $1060\,\mathrm{nm}$.
The primary goal of DES is to study the origin of cosmic acceleration and the
nature of dark energy through four key probes: weak lensing, clustering of the large-scale structure, galaxy clusters, and type Ia supernovae.

The first season of DES observations, from 31$^\textrm{st}$ August 2013 to 9$^\textrm{th}$ February 2014, resulted in the Y1 data stet analysed here \cite{Diehl:2014lea,y1gold}. During Y1, DES observed $\sim1500\,\deg^2$ of the wide-area survey footprint with three to four dithered tilings per filter. The Y1 footprint consisted of two areas: one near the celestial equator including Stripe 82 (S82\cite{2014ApJ...794..120A}), and a much larger area that was also observed by the South Pole Telescope (SPT\cite{2011PASP..123..568C}).
In Fig. 1 of the main text we show with a solid white line the DES Y1 sky coverage in celestial coordinates;
the complete sky coverage for the final survey is shown with a dashed line. Fig. \ref{fig:fermimap} shows the DES Y1 footprint in Galactic coordinates.

In our analysis we make use of the \textsc{metacalibration} shear catalogue \cite{Zuntz:2017pso}. The \textsc{metacalibration} catalogue yields a total of 35 million objects and the final number density of the selection is 5.5 galaxies per square arcminute. 
\textsc{metacalibration} \cite{Huff:2017qxu,Sheldon:2017szh} is a method to estimate weak lensing shear, calibrating associated biases directly from the imaging data, rather than from prior information about galaxy properties or simulations. \textsc{metacalibration} has been shown to be accurate at the part-per-thousand
level in the absence of blending with other galaxies \cite{Sheldon:2017szh}, and at the part-per-hundred level for the blending present in DES Y1 data \cite{Zuntz:2017pso}.

The implementation of \textsc{metacalibration} consists of measuring the two-component ellipticity, ${\bm e}$, of a galaxy in the DES Y1 run by fitting a single Gaussian model to its single-epoch images in the \textit{riz} bands. The galaxy images are then artificially distorted with a known shear, ${\bm\gamma}$, and the ellipticity is measured again to construct the response matrix, ${\bm R}_{\gamma}$, as the derivative of measured ellipticity w.r.t.\ shear. Thus, the ellipticity estimator can be written as the following Taylor expansion \cite{Huff:2017qxu,Sheldon:2017szh,Zuntz:2017pso}:
\be
{\bm e} = {\bm e}|_{\gamma=0} + \frac{\partial \bm e}{\partial {\bm\gamma}}{\bm R}_{\gamma=0}{\bm\gamma} + \ldots
= {\bm e}|_{\gamma=0} + {\bm R}_{\gamma} {\bm\gamma} + \ldots.
\ee
The elements of the response matrix are measured in \textsc{metacalibration} by the numerical derivative:
\begin{align} 
{\bm R}_{\gamma_{i,j}} = \frac{e^{+}_i - e^{-}_i}{\Delta \gamma_j},
\end{align}
where $e^{\pm}_i$ is the measurement of the $i$th ellipticity component made on an
image sheared by $\pm\gamma_j$, and $\Delta \gamma_j = 2 \gamma_j$.

It is also necessary to correct for selection effects, i.e.\ shear biases that may occur when placing a cut on any quantities that change under shear, such as signal-to-noise ratio. This is accomplished by measuring the mean response of the estimator to the selection, repeating the
selections on quantities measured on sheared images. The mean selection response
matrix is given by:
\begin{equation}
\langle {\bm R}_{S_{i,j}} \rangle = \frac{\langle e_i^{S+} \rangle - \langle e_i^{S-} \rangle}{\Delta \gamma_j},
\end{equation}
where $\langle e_i^{S\pm} \rangle$ represent the mean of the ellipticities measured
on images without artificial shearing, but selected by properties measured on sheared images. The full response for the mean shear is given by the sum of the shear response and selection response:
\begin{align}
\langle {\bm R} \rangle = \langle {\bm R}_{\gamma} \rangle + \langle {\bm R}_S \rangle.
\end{align}
When measuring a shear statistics, such as mean shear or a shear two-point function,
the measurement can be appropriately corrected for the mean response $R=\left(\langle\bm{R}\rangle_{11}+\langle\bm{R}\rangle_{22}\right)/2$ to produce a calibrated result.

DES galaxies were assigned to redshift bins using a re-implementation of the \textsc{bpz} algorithm \cite{Benitez2000}, which provides an estimate of the redshift probability density $p_{\rm BPZ}(z_{j})$ of each galaxy $j$. The assignment is done based on fluxes measured by \textsc{metacalibration} on the original and sheared images to correct for selection response of redshift binning. 
The fluxes in the bands \textit{griz} of the \textsc{metacalibration} galaxies
are used for estimating an expectation value of $\langle z \rangle = \int_0^z z_j{\rm d}z_j \, p_{\rm BPZ}(z_{j})$. Thus, four redshift bins are defined by
$0.20 < \langle z \rangle < 0.43$, $0.43 < \langle z \rangle < 0.63$,
$0.63 < \langle z \rangle < 0.90$, and $0.90 < \langle z \rangle < 1.30$. The corresponding four redshift distributions, $n^i(z)$, 
are taken as the stacked $p_{\rm BPZ}(z)$ of the galaxies, estimated from their improved joint-fit photometry \cite{y1gold,Hoyle:2017mee}. The mean values of each redshift bin stacked $p_{\rm BPZ}(z)$ are calibrated independently, and they are found to be consistent with the \textsc{bpz} estimate, with a joint uncertainty of $\sigma_{\langle z\rangle}\approx 0.015$ \cite{Hoyle:2017mee}. 

Since the purpose of this work is detection of the cross-correlation between the UGRB and shear, rather than accurate constraints on cosmological parameters, we do not explicitly account for the
systematic uncertainties in shear and redshift calibration in our analysis. The moderate significance of our measurements and the accuracy of the DES Y1 calibrations justify this choice.

\section{Gamma-ray data} \label{sec:FERMIdata}
\Fermi\ is a \g-ray pair-conversion telescope that has been operating for the last 10 years in space. Due to its wide energy range ($20\,\mathrm{MeV}$ to $1\,\mathrm{TeV}$) and its capability of rejecting the background of charged cosmic rays, it is an excellent instrument to investigate the UGRB. It scans the whole sky every three hours with a remarkable angular resolution
for \g \ rays ($\sim0.1\,\deg$ above $10\,\mathrm{GeV}$).

In this work we used 108 months of data, from mission week 9 to week 476. The photon and exposure maps are produced with the LAT Science Tools version \texttt{v10r0p5} \footnote{
\protect\url{https://fermi.gsfc.nasa.gov/ssc/data/analysis/software/}
}. We selected the Pass8 \textsc{ultracleanveto} class \footnote{See \protect\url{http://www.slac.stanford./glast/groups/canda/lat_Performance.htm}
}, which has the lowest cosmic-ray contamination and is the most appropriate class for diffuse emission analysis. The Fermi Tools provide the possibility of choosing different angular resolutions, which are organised in four quartiles, from PSF0 to PSF3, corresponding to a transition from the worst to the best PSF. In order to have a balance between the photon count statistics and a good direction reconstruction, we selected the best quartile PSF3 for energies below $1.2\,\mathrm{GeV}$ (where we have the highest photon counts) and PSF1+2+3 for higher energies. The PSF is modelled according the \Fermi\ specifications \footnote{ \protect\url{https://fermi.gsfc.nasa.gov/ssc/data/analysis/documentation/Cicerone/Cicerone_LAT_IRFs/index.html}
} and for each energy bin in our analysis an effective PSF is determined by weighting the energy-dependent PSF by the intensity energy spectrum of the UGRB. 

We produced 100 intensity maps in \texttt{HEALPix} projection with $N_{\rm side}=1024$, evenly
spaced in logarithmic scale between $100\,\mathrm{MeV}$ and $1\,\mathrm{TeV}$, by dividing the count maps by the exposure and the pixel area. The size of such energy bins is small enough that the exposure can be approximated by its mean value within the energy bin when deriving the flux. The resulting flux maps are then re-binned into 9 larger energy bins between $631\,\mathrm{MeV}$ to $1\,\mathrm{TeV}$ by simply adding up fluxes from the smaller bins. We discard very low energies because the angular resolution is too poor for our purposes.

Since we are interested only in the UGRB, we benefit from excluding Galactic emission and resolved point sources. This is achieved by a process of masking and subtracting described below.

\subsection{Masking \g-ray data}
We build a set of masks according to the following two criteria:
\begin{enumerate}
\item Low latitudes, where the Galactic foreground is stronger, are removed by a flat cut excluding the region between $\pm30\,\deg$ of latitude.
\item Sources identified in the list FL8Y are masked. FLY8 \footnote{ \protect\url{https://fermi.gsfc.nasa.gov/ssc/data/access/lat/fl8y/}
} has been recently released by the \Fermi\ Collaboration as a preliminary version of the upcoming 4FGL catalogue. It contains 5523 sources. Above 10 GeV, we mask also the sources which are present in the 3FHL catalogue \cite{AjeloEtAll2017}, that is more accurate for high energy sources. Each source is masked taking into account both its source brightness and the detector PSF resolution in the specific energy bin. The masking radius $R$ has been defined by:
\be
F_{\Delta E}^g \, \exp{\left(-\frac{R^2}{2 \theta_{\Delta E}^2}\right)} > \frac{F_{\Delta E,\rm faintest}^g}{5}
\label{eq:radius}
\ee
where $F_{\Delta E}^g$ is the integral flux of the source in a given energy bin
$\Delta E$, $F_{\Delta E,\rm faintest}^g$ is the flux of the faintest source in
the same energy bin (and once divided by 5 provides an approximate estimate of the noise, i.e.,  the faintest source emission is approximately a measure of the $5\sigma$ level), and $\theta_{\Delta E}$ is the 68\%
containment angle in that energy bin, as provided by the \Fermi\ PSF. We verified that the non-Gaussian tail of the PSF (not included in the Gaussian approximation in Eq.~\eqref{eq:radius}) does not appreciably contaminate our maps.
\end{enumerate}
This strategy aims at masking the Galactic plane and resolved sources over a sufficiently large area, in order to reduce the chance to have artefacts in the APS produced by source leakage and foreground emission. For further details and impact of the mask, see also ref. \cite{Ackermann:2018wlo}.

\begin{figure*}[t]
    \centering
    \includegraphics[width=0.49\textwidth]{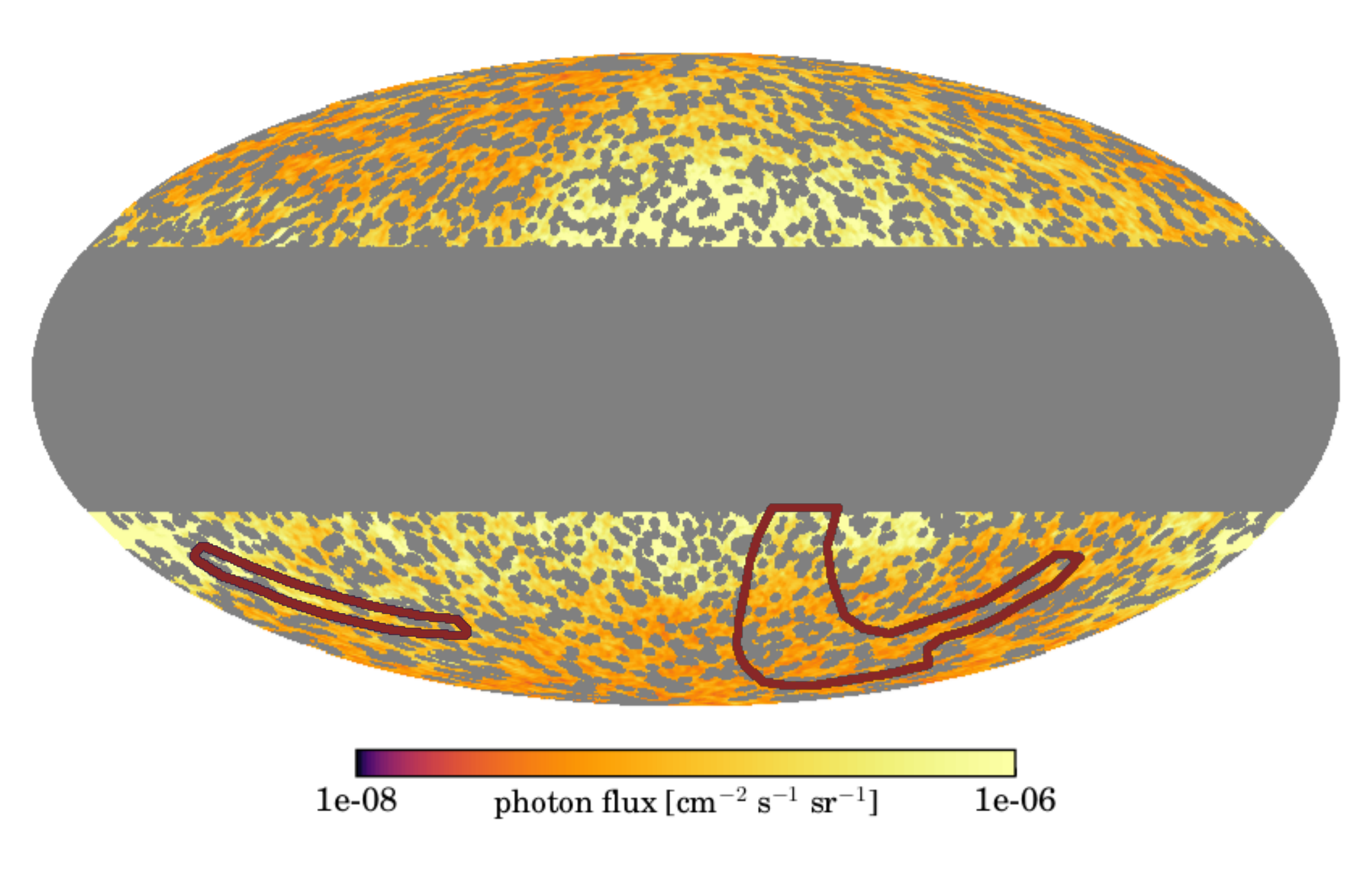}
    \includegraphics[width=0.49\textwidth]{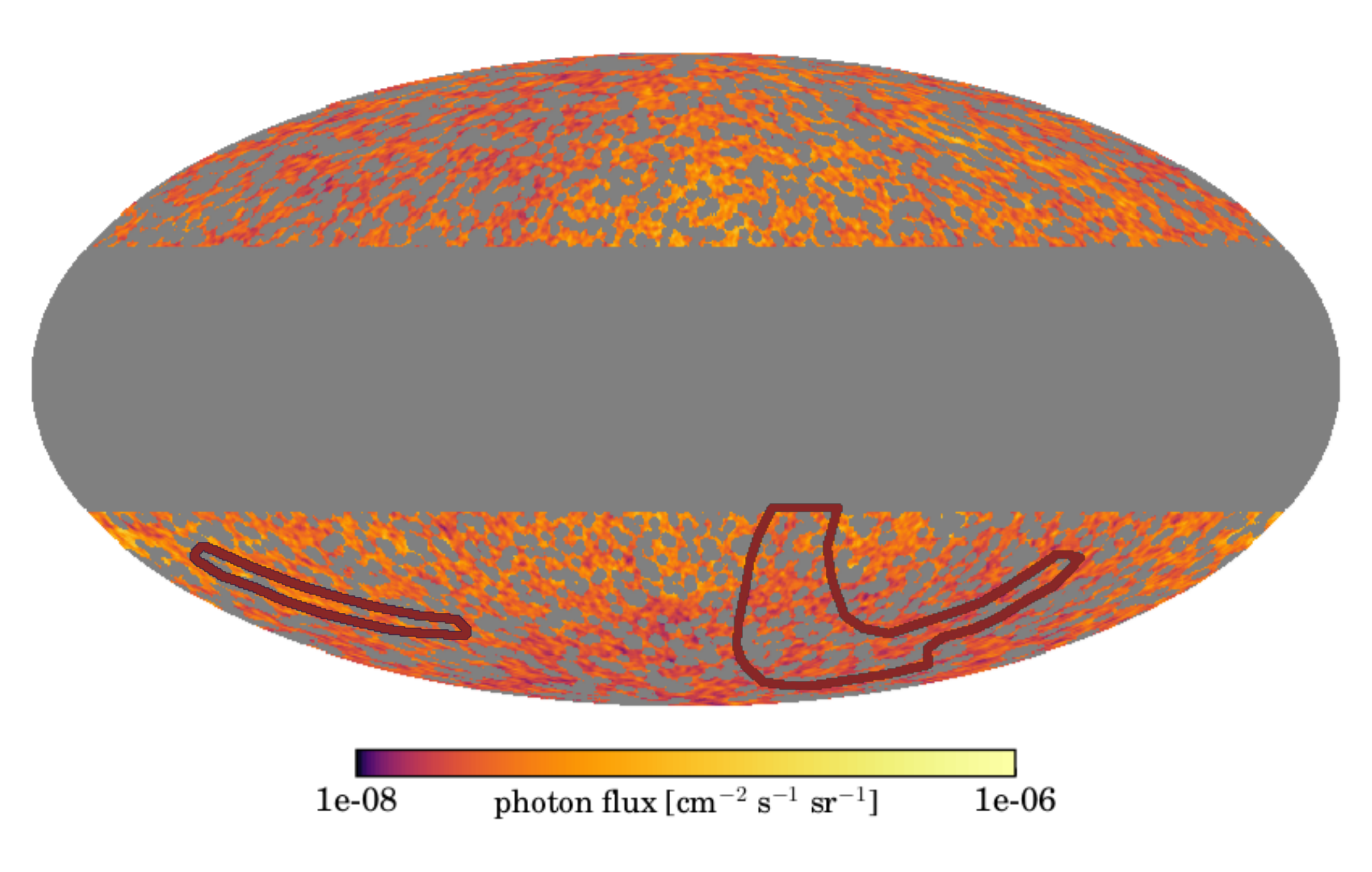}
    \caption{Masked \g-ray map in the $(1.2-2.3)$ GeV energy bin before (left) and after (right)
		the foreground subtraction. The maps have been downsized to $N_{\rm side} = 128$
		and smoothed with a Gaussian beam of size $\sigma = 0.4^\circ$ for illustration purposes. The DES Y1 sky footprint is shown with red solid line.}
    \label{fig:fermimap}
\end{figure*}

\subsection{Foreground removal}
Despite the majority of the Galactic foreground has been removed by cutting out the Galactic plane, still high-latitudes might contain some contamination that causes noise in our analysis. While galactic emission does not contribute to the cross-correlation signal with gravitational shear, nevertheless a foreground contamination adds to the error budget. We therefore performed a galactic foreground subtraction by adopting the template maps provided by the \Fermi\ Collaboration with the Galactic emission model {\tt gll\_iem\_v06.fits} \footnote{ \protect\url{https://fermi.gsfc.nasa.gov/ssc/data/access/lat/BackgroundModels.html}
}.
The foreground template is projected in HEALPix maps with the same $N_{side}$ as the intensity maps and in the same 100 energy bins.
Each template map is convolved with the \Fermi\ PSF and assigned a free normalization. This component is added to a
free constant (representing the UGRB and cosmic-ray contamination) and a Poissonian
likelihood fit is performed globally on all the masked intensity maps.
All obtained best-fit normalization parameters are of the order of unity,
supporting a successful description of the foreground emission.
The normalised foreground templates are then re-binned into the 9 larger energy
bins and subtracted from the corresponding intensity maps. 

In Fig.~\ref{fig:fermimap}, we show an example of the \Fermi\ \g-ray intensity map in the (1.2-2.3) GeV energy bin with the application of the mask described above and illustrating the effect of the Galactic foreground subtraction.

\begin{figure*}[t]
  \centering
  \includegraphics[width=0.55\textwidth]{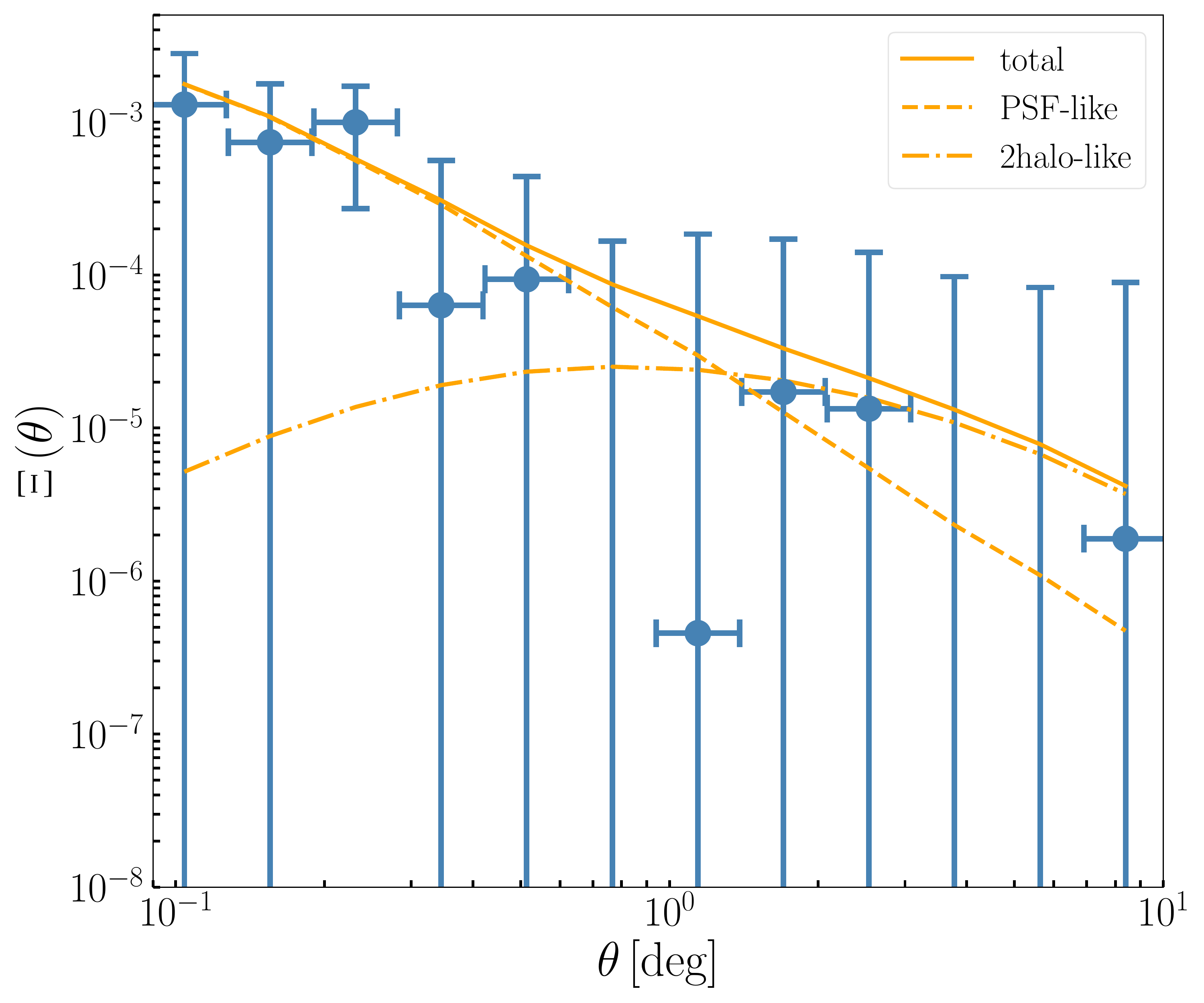}
  \caption{Same as Fig.~2 of the main text but without foreground removal (and normalized as in Fig.~2 of the main text, being $\Xi$ dimensionless).}
  \label{fig:foreg}
\end{figure*}

The impact of foreground removal in cross-correlation studies involving \g-rays has been discussed in Refs. \cite{Xia:2015wka,Shirasaki2014,Shirasaki:2016kol,Ammazzalorso:2018evf,Shirasaki:2018dkz}, where it was shown that the effect is marginal. As mentioned, the main effect of foreground residuals on our study would be a mild reduction of detection significance. This can be appreciated in Fig.~\ref{fig:foreg}, where we show the analogous of Fig.~2 of the main text, but without foreground removal. Since the amplitude $\Xi$ is dimensionless, in Fig. \ref{fig:foreg} it has been normalized as in Fig.~2 of the main text (i.e. relative to the mean intensity $\langle I_\gamma \rangle $ of the foreground-subracted maps), to allow a direct comparison of the two plots. Put in different words,Supplemental Material

\section{Estimation of the covariance matrix} \label{sec:cov}
The covariance matrix is estimated 
by combining the theoretical Gaussian large-scale structure covariance with realizations of the shape-noise term generated from the data. The latter term is expected to be dominating in the covariance, while the former should be small. To avoid notation conflict between the symbol \g\ usually employed for both the weak lensing effect of gravitational shear, and for \g-rays, we shall here simply refer to shear through indexes $r,s$, labelling redshift bins, and to the UGRB anisotropies via indexes $a,b$, labelling energy bins.

In the Gaussian approximation, the generic element of the theoretical harmonic-space covariance matrix $\widehat{\bm\Gamma}$ reads:
\be
\widehat\Gamma_{ar\ell,bs\ell^\prime} = \frac{ \delta^{\rm K}_{\ell\ell^\prime}}{(2\ell+1)\Delta \ell f_{\rm sky} }\left[C_\ell^{ar}C_{\ell^\prime}^{bs}+ \big(C_{\ell^\prime}^{rs}+ \nn^{rs}\big) \big(C_\ell^{ab} + \nn^{ab}\big)\right].\label{eq:covCl}
\ee
It represents the covariance between the measurements of two cross-correlation harmonic-space power spectra: one cross-correlating \g-ray map in energy bin $a$ with shear map in redshift bin $r$; and another considering \g-ray map in energy bin $b$ with shear map in redshift bin $s$. In the Gaussian approximation, these two cross-correlation power spectra, $C^{ar}_\ell$ and $C^{bs}_{\ell^\prime}$, have a covariance which is diagonal in $\ell\ell^\prime$, and which depends on both the cross-correlation signals, as well as the \g-ray--\g-ray and shear-shear auto-correlations, i.e.\ $C^{ab}_\ell$ and $C^{rs}_{\ell^\prime}$. 
Both auto and cross-correlation theoretical signals involving \g-rays have been corrected for the effect of the {\it Fermi}-LAT PSF beam function.
In Eq.~\eqref{eq:covCl}, $\nn$ denotes the noise terms. Note that the noise does not depend on the angular scale. In Eq. (\ref{eq:covCl}) $f_{\rm sky}$ accounts for the incomplete sky coverage, with $f_{\rm sky}^{\rm DES} = 0.042$ denoting the DES footprint (independent of the redshift bin) and $f_{\rm sky}^a = (0.315,0.289,0.404,0.449,0.468,0.475,0.476,0.477,0.477) $ being the fraction of unmasked pixels of the $\gamma$-ray map in the $a$-th energy bin. For the cross-correlation estimator with complex masks as we have here, an unambiguous definition of $f_{\rm sky}$ is not possible. We tested two options: a geometric mean $f_{\rm sky} = (f_{\rm sky}^a f_{\rm sky}^{\rm DES})^{1/2}$ and the $f_{\rm sky}$ relative to the overlap of the DES footprint with the {\it Fermi}-LAT unmasked sky. We comment more in the two alternatives below, but we anticipate that results on the SNR arising from the two choices exhibit only a minor change. 

For details on the Gaussian covariance matrix, see Section 5.1 in ref. \cite{Camera:2014rja}.
Since our measurements are performed in real space, we compute the Legendre transform of Eq.~\eqref{eq:covCl} to return the real-space covariance matrix, $\bm\Gamma$, with entries $\Gamma_{ar\vartheta,bs\vartheta^\prime}$.

As mentioned above, the dominant contribution to the covariance is represented by the shape noise term. In the Gaussian approximation, it takes the form of $\nn^{rs} (C_\ell^{ab}+\nn^{ab})$. In order to estimate this contribution more accurately, without resorting to  theoretical approximations, we produce 2000 realisations of the noise directly from the data in real space. To this end, we rotate each galaxy in our catalogue by an independent random angle. The shear signal as measured from these rotated source catalogues represents a random realisation of shape noise (see e.g. refs. \cite{Gruen:2017xjj,Hikage:2018qbn,Murata:2017zdo}).

We complete the construction of the total covariance by adding a theoretical estimate of the large-scale structure term, $C_\ell^{ar}C_\ell^{bs}+C_\ell^{rs} (C_\ell^{ab} + \nn^{ab})$. We assume it subdominant in the covariance budget, and this is verified a posteriori. The \g-ray auto-correlation $C_\ell^{ab}$ entering in the theoretical estimate can be well-fitted by a simple model, given by a power-law plus a constant, i.e.\ $C_{\ell{\rm ,mod}}^{ab} = A_{ab}\ell ^{-\alpha_{ab}}+C_{\rm P}^{ab}$ (see e.g. ref. \cite{Ackermann:2018wlo}). We fit the three parameters $\{A_{ab},\,\alpha_{ab},\,C_{\rm P}^{ab}\}$ for each energy bin pair against the measurement of the auto-correlation power spectrum obtained by running the tool \texttt{PolSpice} (as in ref. \cite{Ackermann:2018wlo}). Finally, the \g-ray noise term, $\nn^{ab}$, is computed with the analytic estimator of eq. 5 in ref. \cite{Fornasa:2016ohl}.

The shear auto-correlation is derived using the galaxy redshift distributions described above, and adopting $\Lambda$CDM cosmology with parameter fiducial values from \textit{Planck} \cite{Planck2015}. The model is compatible with previous measurements \cite{Troxel2018}. 

The large-scale structure part of the covariance is added to the shape noise term by using the following procedure. We create a set of 2000 simulated datasets from a multivariate Gaussian distribution with zero mean and covariance equal to the large-scale structure part of the theoretical covariance. Then, we add this simulated data to the shape noise realisations discussed above, thus obtaining 2000 samples containing both terms. These 2000 samples are then used to obtain the covariance matrix. The inverse of the covariance estimated such is a biased estimator of the inverse covariance, with the bias depending on the number of realisations, number of bins, and parameters of the model chosen to fit the data  \cite{anderson2009,Hartlap2007}. We apply the Anderson-Hartlap correction in order to de-bias the inverse covariance (as done e.g.\ in Section 3 of ref. \cite{Kilbinger2013}).

\begin{figure*}[t]
    \centering
    \includegraphics[width=0.49\textwidth]{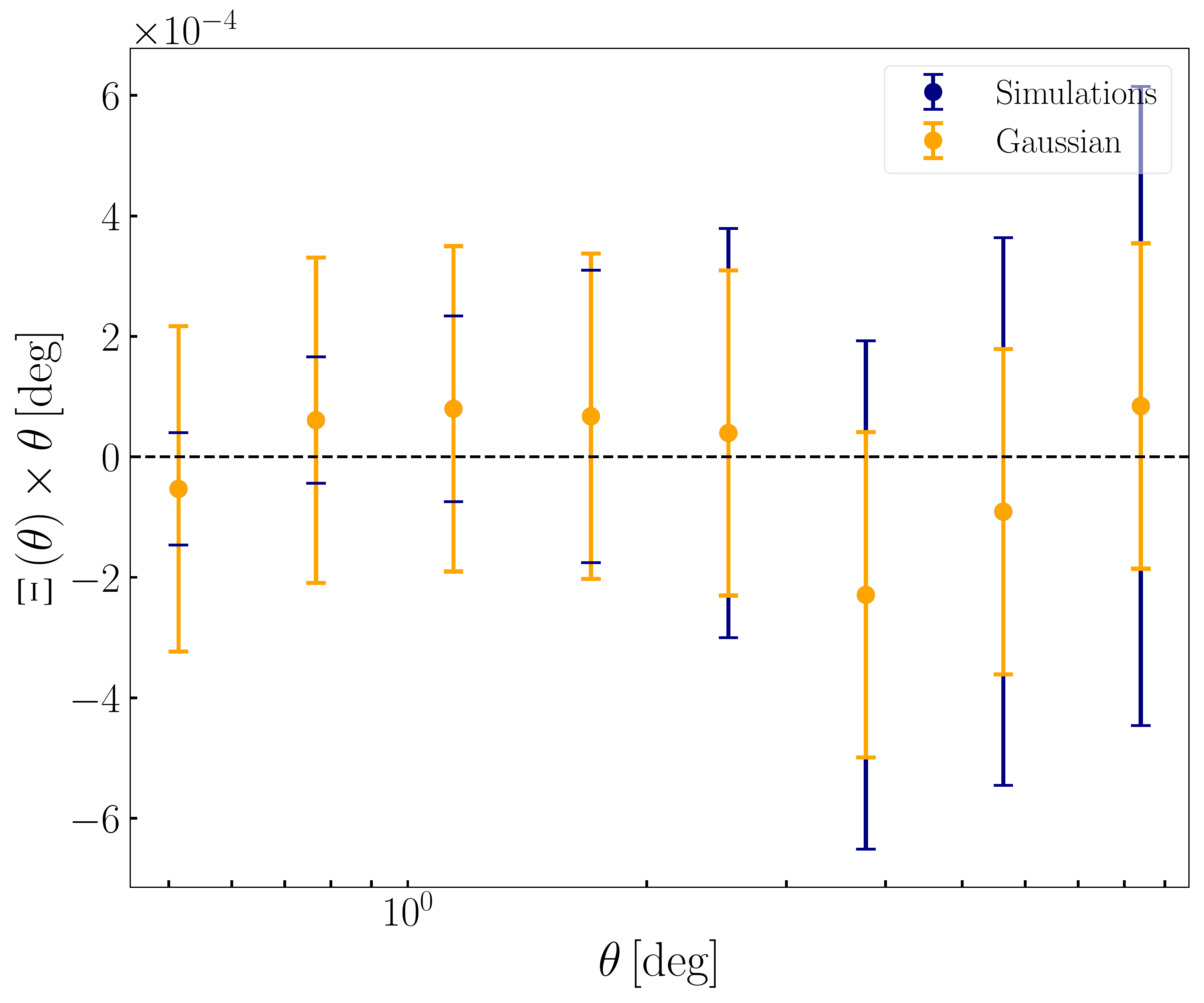}
    \includegraphics[width=0.49\textwidth]{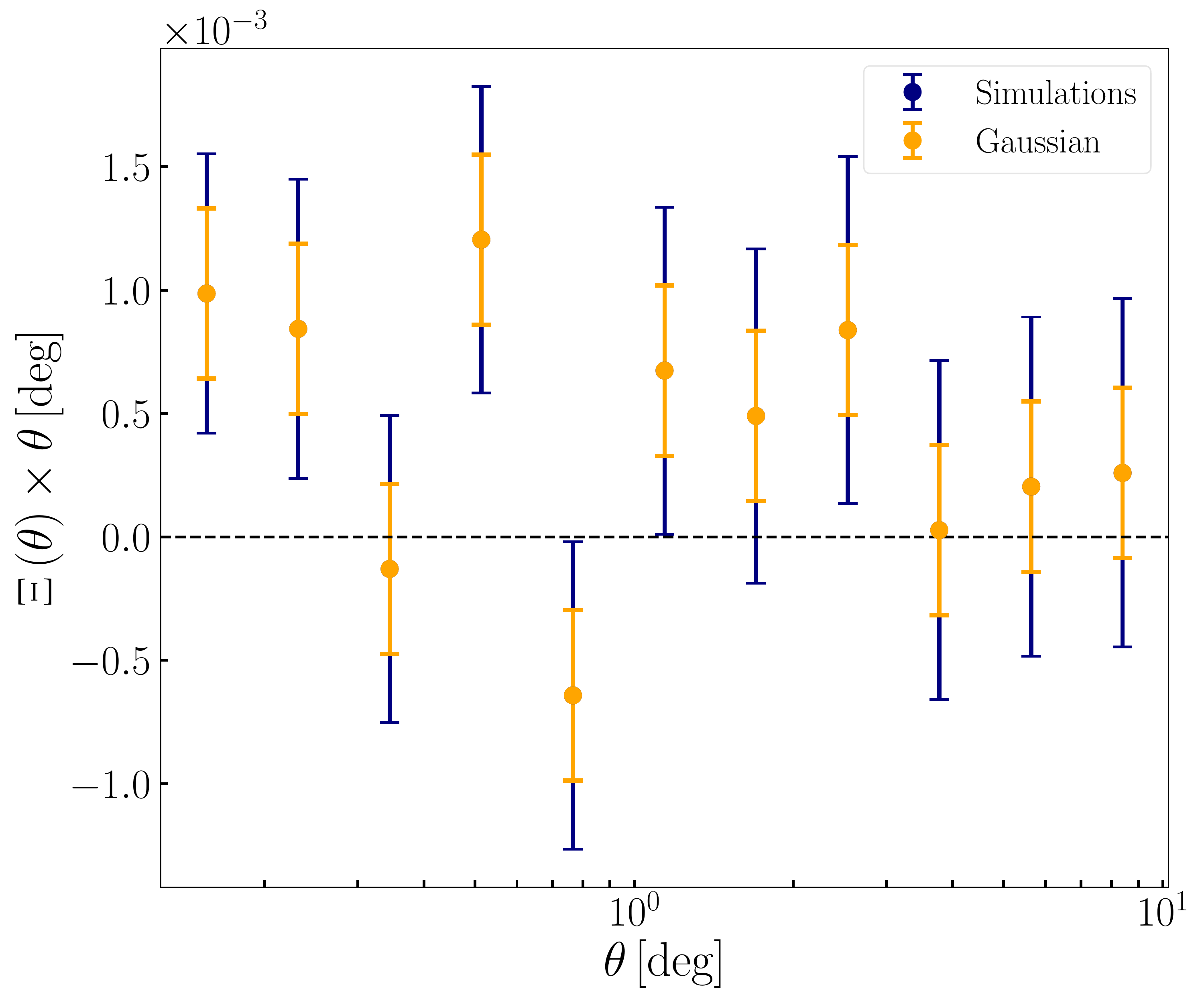}
    \caption{Comparison between the shape-noise standard deviation computed with the theoretical Gaussian approximation (yellow error bars) and via the simulations (blue error bars) for two energy bins in the first DES redshift bin. We focus only on angular scales above the \Fermi\ PSF in the specific energy bin. \textit{Left panel:} energy bin between $1.2$ and $2.3\,\mathrm{GeV}$. \textit{Right panel:} energy bin between $131$ and $1000\,\mathrm{GeV}$.}
    \label{fig:varcomp}
\end{figure*}

In order to validate our procedure, we compare the shape noise term obtained via the simulations with the analogous term from the theoretical estimate, $\nn^{rs} (C_\ell^{ab} + \nn^{ab})$. In Fig.~\ref{fig:varcomp}, we show a comparison between the theoretical shape noise standard deviation and the results from the simulations, in the case of a low-energy bin (left panel) and a high-energy bin (right panel) 
combined with the lowest DES redshift bin.
We see that the variance obtained with the two techniques approximately agrees, yet can differ due to complex masking effects present in the data \cite{Troxel2018b}. When deviations are present, the variance obtained with simulations is typically larger, as expected. By adopting the shape noise term directly derived from the data, we ensure that the quoted errors and values for the goodness of fit are correct. 

We compare also the large-scale structure term with the covariance matrix obtained from simulations after rescaling the latter by $C_\ell^{rs}/\nn^{rs}$. Results are in very good agreement with the choice of the effective parameter $f_{\rm sky}$ entering Eq. (\ref{eq:covCl}) as the geometric mean of DES and \Fermi\ sky coverage. Another value for the $f_{\rm sky}$ parameter that is frequently used in the literature is the overlap between the two masks. We verified that, in this case, results would be just slightly modified, with the SNR becoming 4.8 instead of 5.3, as obtained in the main analysis.

\section{Theoretical models} \label{sec:models}
The harmonic-space (often referred to as `angular') power spectrum of the cross-correlation between a map of $\gamma$-rays in the $a$th energy bin and a map of gravitational shear in the $r$th redshift bin can be computed as:
\be
C_\ell^{ar} = \int\de E\, \de z\,\frac{1}{H(z)}\frac{W_\textrm{\rm gamma}^a(E,z)W_{\rm shear}^r(z)}{\chi(z)^2}
P_{\gamma\delta}\!\!\left[k=\frac{\ell}{\chi(z)},z\right],
\label{eq:clgen}
\ee
where $E$ is the \g-ray energy, and $\chi(z)$ is the comoving distance to redshift $z$, obeying $\de z/\de\chi=H(z)$ with $H(z)$ the Hubble rate. Formally, the integrals extend over the whole \g-ray energy spectrum and from $z=0$ to the horizon, but the window functions $W_\textrm{\rm gamma}^a(E,z)$ and $W_{\rm shear}^r(z)$ implement energy and redshift binning effectively reducing the integration range (for details, see next Subsections). Finally, $P_{\gamma\delta}$ is the three-dimensional cross-power spectrum between a given \g-ray population sourcing the UGRB emission and the matter density contrast $\delta$. It represents the three-dimensional correlation (in Fourier space) between what seeds the unresolved \g-ray emission and what sources the gravitational lensing effect, namely matter inhomogeneities. It is a function of both redshift and physical scale $k$, the modulus of the physical wavenumber. In the Limber approximation, $k$ and the angular multipole $\ell$ are linked by $k=\ell/\chi(z)$. This approximation is valid for $\ell\gg1$, which is the case of the present work.

As mentioned before, the quantity measured from the data is the real-space cross-correlation of UGRB anisotropies
with tangential shear, which
is related to the harmonic-space cross-power spectrum of Eq.~\eqref{eq:clgen} by a Legendre transform:
\be
\hat\Xi^{ar}(\theta) = \sum_\ell \frac{2\ell+1}{4\pi\ell(\ell+1)} C_\ell^{ar}\, W_\ell^a\,P^{(2)}_\ell(\cos\theta),
\label{equ:Cl2xi}
\ee
with $\theta$ being the angular separation on the sky, $P^{(2)}_\ell$ the Legendre polynomial of order two, and $W_\ell^a$ the {\it Fermi}-LAT PSF beam function in energy bin $a$, accounting for the finite resolution of the detector.

\subsection{Gravitational lensing window function}
\label{sec:wf1}
The window function describes the mean distribution of the signal along the line of sight, in the given energy or redshift bin.
For the gravitational lensing effect, the window function is given by (see e.g. ref. \cite{Bartelmann2010a}):
\be
	W_{\rm shear}^r(z) =
	\frac{3}{2} H_0^2 \om (1+z)\chi(z) \int_{z}^\infty \de z^\prime\, \frac{\chi(z^\prime)-\chi(z)}{\chi(z^\prime)} n^r(z^\prime),
	\label{eq:W_lens}
\ee
where $H_0\equiv H(z=0)$ is the Hubble constant, $\om$ is the matter abundance in
the Universe (sum of the dark matter and the baryon abundances, $\odm$ and $\ob$), and $n^r(z)$ is the redshift distribution of background galaxies in
the lensing data set in bin $r$. The galaxy distribution depends on the data set and redshift cut, as described above.

\subsection{WIMP-sourced \g-rays window function}
\label{sec:wf2}
The window function for UGRB anisotropies sourced by annihilations of \DM\ particles reads \cite{Ando2006,Fornengo:2014}:
\be
	W_\textrm{{\rm gamma},DM}^a(E,z) = \frac{(\odm \rho_{\rm c})^2}{4\pi} \frac{\sv}{2\mdm^2} \left(1+z\right)^3 \Delta^2(z) \frac{\de N_{\rm ann}}{\de E}\left[E(1+z) \right] e^{-\tau\left[E(1+z),z\right]} \ ,
	\label{eq:win_annDM}
\ee
where $\rho_{\rm c}$ is the critical density of the Universe, $\mdm$ is the mass of the \DM\ particle, and $\sv$ denotes the velocity-averaged annihilation cross-section, assumed here to be the same in all dark matter haloes. 

Among the other ingredients, we have $\de N_{\rm ann} / \de E$, indicating the number of photons produced per annihilation as a function of energy, and setting the \g-ray energy spectrum. We will consider it to be given by the sum of two contributions: prompt \g-ray production from \DM\ annihilations (or decays); and inverse Compton scattering of \DM\ produced electrons and positrons on CMB photons (we compute inverse Compton assuming negligible magnetic field and no diffusion for the produced electrons and positrons). Results of our analysis will be shown for two annihilation final states: $b\bar b$ pairs, which yield a relatively soft spectrum of photons and electrons, mostly associated to hadronisation into pions and their subsequent decay; $\tau^+\tau^-$, which provides a harder spectrum, mostly associated to final state radiation of photons and decay of the muons produced by $\tau$ decay into electrons (with subsequent \g-ray emission through inverse Compton on CMB), with an additional semi-hadronic decay into pions \cite{Fornengo2004,Cembranos2011,Cirelli2011}.

Then, the optical depth $\tau$ in Eq.~\eqref{eq:win_annDM} accounts for attenuation of \g-rays by the extra-galactic background light, and is taken from the literature\cite{Finke2010}. Finally, the clumping factor $\Delta^2$ is related to the fact that the signal of annihilating \DM\ depends on the average of the square of the \DM\ density; it is defined as (see e.g. ref. \cite{Fornengo:2014}):
\be
\label{eq:clumping}
	\Delta^2(z) \equiv
	\frac{\langle \rho^2_{\rm DM} \rangle}{{\bar \rho}^2_{\rm DM}} =
	\int_{M_{\rm min}}^{M_{\rm max}} \de M \frac{\de n_{\rm h}}{\de M}(M,z) \,\left[1+b_{\rm sub}(M,z)\right]
	\times \int \de^3 x\, \frac{\rho^2_{\rm h}({\bm x}|M,z)}{{\bar \rho}^2_{\rm DM}},
\ee
where: $\de n_{\rm h}/\de M$ is the halo mass function \cite{Sheth1999}; $M_{\rm min}$ is the minimal halo mass, which we fix to $10^{-6} M_\odot$; $M_{\rm max}$ is the maximal mass of haloes, for which we use $10^{18} M_\odot$, although results are insensitive to the precise value assumed; $\rho_{\rm h}({\bm x}|M,z)$ is the \DM\ density profile of a halo with mass $M$ at redshift $z$, taken to follow a Navarro-Frenk-White profile \cite{Navarro1997}; and $b_{\rm sub}$ encodes the `boost' to the halo emission provided by subhaloes. To characterise the halo profile and the subhalo contribution, we need to specify their mass concentration. For the main haloes we follow ref. \cite{Prada2012}. On the other hand, the description of the concentration parameter $c(M,z)$ at small masses and for subhaloes is still an open issue and provides a source of uncertainty. We considered the two most updated analyses, from Ref. \cite{Moline2016}, where we assume $\de n_{\rm sub}/ \de M_{\rm sub}\propto M_{\rm sub}^{-2}$ (see also Ref. \cite{Sanchez-Conde2014}), and from Ref. \cite{Hiroshima:2018kfv}. We show only the former, since the two descriptions just differ by roughly a constant factor (around 10) in the derived final value of the annihilation cross section.

\subsubsection{Astrophysical \g-ray sources window function}\label{sec:wf3}

Besides the possible particle \DM\ emission, \g-rays in the UGRB are certainly produced by a number of astrophysical sources. The most relevant \g-ray emitters include: blazars, misaligned active
galactic nuclei, and star forming galaxies. Their window function is given by:
\be
    W_\textrm{{\rm gamma},S}^a(E,z) = \chi(z)^2 \int_{\mathcal{L}_{\rm min}}^{\mathcal{L}_{\rm max}(F_{\rm sens},z)} \de \mathcal{L}
   \, \Phi_{\rm S}(\mathcal{L},z,E) \, \frac{\de N_{\rm S}}{\de E}\left(\mathcal{L},z\right)\times e^{-\tau\left[E(1+z),z\right]}  \ ,
    \label{eq:win_astro}
\ee
where $\mathcal{L}$ is the \g-ray rest-frame luminosity in the energy interval $0.1$ to $100\,\mathrm{GeV}$, $\Phi_{\rm S}$ is the \g-ray luminosity function (GLF) of the source class S of astrophysical emitters included in our analysis, and $\de N_{\rm S}/\de E$ is its observed (unabsorbed) energy spectrum. The upper bound, $\mathcal{L}_{\rm max}(F_{\rm sens},z)$, is the luminosity above
which an object is resolved in the FL8Y and 3FHL catalogues, and consequently masked in our analysis. As we are interested in the contribution from unresolved astrophysical sources, only sources with luminosity smaller than $\mathcal{L}_{\rm max}$ are included. Conversely, the minimum luminosity, $\mathcal{L}_{\rm min}$, depends on the properties of the source class under consideration. 

We consider a unified blazar model combining BL Lacertae and flat-spectrum radio quasars as a single source class. The GLF and energy spectrum are taken from ref. \cite{Ajello2015} where they are derived from a fit to the properties of resolved blazars in the third \Fermi\ catalogue \cite{Acero2015}. In the case of misaligned AGNs, we follow ref. \cite{Di-Mauro2013}, who built the GLF from the radio luminosity function of misaligned AGNs. We consider their best-fitting relation between the \g-ray and radio luminosities $\mathcal{L}$ -$L_{r,{\rm core}}$ and assume a power-law spectrum with index $\alpha_{\rm mAGN}=2.37$. To derive the GLF of star-forming galaxies, we start from the infrared luminosity function \cite{Gruppioni2013} (adding up spiral, starburst, and SF-AGN populations of their Table~8). Then we relate \g-ray and infrared luminosities using the best-fitting $\mathcal{L}$-$L_{\rm IR}$ relation from ref. \cite{Ackermann2012}. The energy spectrum is taken to be a power-law with spectral index $\alpha_{\rm SFG}=2.7$.

\subsection{Three-dimensional seed power spectra}
\label{sec:ps}

To compute the three-dimensional cross-power spectrum $P_{\gamma\delta}$ between the clustering of a given population of \g-ray emitters and the matter density field, we follow the halo model formalism (e.g. ref. \cite{Cooray2002}), and write $P_{\gamma\delta}=P_{\gamma\delta}^{\rm 1h}+P_{\gamma\delta}^{\rm 2h}$. Below we derive the 1- and 2-halo terms for the various cases (see also ref. \cite{Fornengo:2014}).

\subsubsection{Dark matter \g-ray sources}
The 3D cross power spectrum between \g-ray emission from particle \DM\ and matter density is given by:
\begin{align}
    P_{\gamma_{\rm DM}\delta}^{\rm 1h}(k,z) &= \int_{M_{\rm min}}^{M_{\rm max}} \de M\ \frac{\de n_{\rm h}}{\de M}(M,z) \,\hat v_{\gamma_{\rm DM}}(k|M,z) \, \hat u_\delta(k|M,z) \\
    P_{\gamma_{\rm DM}\delta}^{\rm 2h}(k,z) &= \left[\int_{M_{\rm min}}^{M_{\rm max}} \de M\,\frac{\de n_{\rm h}}{\de M}(M,z)\, b_{\rm h}(M,z) \,\hat v_{\gamma_{\rm DM}}(k|M,z) \right] \left[\int_{M_{\rm min}}^{M_{\rm max}} \de M\,\frac{\de n_{\rm h}}{\de M}(M,z)\,b_{\rm h}(M,z) \hat u_\delta(k|M,z) \right]P^{\rm lin}(k,z),
\label{eq:PSDM}
\end{align}
where $P^{\rm lin}$ is the linear matter power spectrum, $b_{\rm h}$ is the linear bias (taken from the model of ref. \cite{Sheth1999}), and $\hat u_\delta(k|M,z)$ is the Fourier transform of the matter halo density profile, i.e.\ $\rho_{\rm h}({\bm x}|M,z)/\bar \rho_{\rm DM}$. The Fourier transform of the \g-ray emission profile from annihilating \DM\ is described by $\hat v_{\gamma_{\rm DM}}(k|M,z)$, and it is related to the square of the \DM\ density profile. For its precise form, see the appendix of ref. \cite{Cuoco2015}.

\subsubsection{Astrophysical \g-ray sources}
The cross-correlation of the matter density with astrophysical \g-ray sources is given by the 3D power spectrum:
\begin{align}
     P_{\gamma_{\rm S}\delta}^{\rm 1h}(k,z) &= \int_{\mathcal{L}_{\rm min}}^{\mathcal{L}_{\rm max}} \de \mathcal{L}\,\frac{\Phi_{\rm S}(\mathcal{L},z)}{\langle f_{\rm S} \rangle} \frac{\de F}{\de E}\left(\mathcal{L},z\right) \hat u_\delta\left[k|M(\mathcal{L},z),z\right]	\label{eq:PSastro1h}\\
     P_{\gamma_{\rm S}\delta}^{\rm 2h}(k,z) &= \left[\int_{\mathcal{L}_{\rm min}}^{\mathcal{L}_{\rm max}} \de \mathcal{L}\, b_{\rm S}(\mathcal{L},z)\,\frac{\Phi_{\rm S}(\mathcal{L},z)}{\langle f_{\rm S} \rangle} \frac{\de F}{\de E}\left(\mathcal{L},z\right) \right]
      \left[\int_{M_{\rm min}}^{M_{\rm max}} \de M\,\frac{\de n}{\de M} b_{\rm h}(M,z) \hat u_\delta(k|M,z) \right] 
      P^{\rm lin}(k,z) ,
	\label{eq:PSastro}
\end{align}
where $b_{\rm S}$ is the bias of \g-ray astrophysical sources with respect to the matter density, for which we adopt $b_{\rm S}(\mathcal{L},z)=b_{\rm h}[M(\mathcal{L},z)]$. That is, a source with luminosity $\mathcal{L}$ has the same bias $b_{\rm h}$ as a halo with mass $M$, with the relation $M(\mathcal{L},z)$ between the mass of the host halo $M$ and the luminosity of the hosted object $\mathcal{L}$ taken from ref. \cite{Camera:2014rja}. The mean flux $\langle f_{\rm S} \rangle$ is defined as $\langle f_{\rm S} \rangle=\int \de \mathcal{L}\de F/\de E \Phi_{\rm S}$.

\section{Blinding and unbliding procedure} \label{sec:blind}

The analysis has been performed by adopting a blinding technique, which relied on the execution of the cross-correlation analysis on nine combinations of data -- one true and the other artificial. For both $\gamma$-rays and gravitational lensing we adopted 3 data variants, one of which was the true one.
After the data were created, they were randomly assigned symbolic names. The analysis was then performed on all 9 combinations of data, without knowing which was the true version. 

The aim of the blinding 
was to ensure that our analysis would not falsely detect a signal. Criteria to test the blind analysis were defined beforehand. After recording the results of the blind analysis and agreeing on its interpretation, the symbolic name assignments were revealed.

The construction of the versions of the data and the unblinding procedure is discussed below.

\subsection{Blinded \g-ray maps}
The $\gamma$-ray maps adopted in the blind analysis are:
\bi
\item G0: The true {\it Fermi}-LAT $\gamma$-ray maps.
\item G1: Poissonian random map with a constant expected photon count over all the unmasked pixels, i.e.~in each pixel of the map we extract a random number from a Poissonian distribution with fixed mean. The mean was computed by taking the average counts of the real maps in the unmasked pixels in each energy bin, and then multiplying it by a factor of 10 in order to simulate improved statistics. The produced counts maps are then transformed into flux maps by the usual procedure of dividing them by the mean detector exposure in each energy bin and by the pixel area.
\item G2: Random reshuffle of all unmasked pixels of the real map in each energy bin.
\ei

Once produced, the three sets of maps have been blindly and randomly assigned names A, B and C and the association stored.

\subsection{Blinded shear maps}
The shear signals adopted in the blind analysis are:

\begin{itemize}
\item K0: The true DES tangential component of galaxy shapes (this is the shear component for which we are looking for the cross-correlation  signal when combined with the {\it Fermi} maps).
\item K1: The cross-component (also known as $B$-modes or $\gamma_\times$) of galaxy shapes (which instead should yield pure shape noise and a null detection when correlated with the {\it Fermi} maps).
\item K2: A linear combination of the null signal given by K1 and the cross-correlation signal of \textsc{redMaGiC} galaxies at redshift $z=0.2 - 0.45$ \cite{Prat2018}, with the linear combination coefficients chosen such that the signal should be neither plainly visible nor certainly undetectable. Since the correlation between \textsc{redMaGiC} galaxies and gravitational shear is significant, this mock data set is meant to
inject in our analysis a situation potentially (but not necessarily) leading to what could be seen as a detection.
\end{itemize}

Once produced, the three versions of the data have been blindly and randomly assigned names X, Y and Z and the association stored.
We note here that the case K1 provides the cross-shear null test investigated in previous attempts of measurement of the signal~\cite{Shirasaki2014,Shirasaki:2016kol,Troster:2016sgf,Shirasaki:2018dkz}.

\subsection{Blind analysis and unblinding procedure}

The nine versions of cross-correlations results were all processed and vetted, with no team member aware of which combination represented the true data vector. We proceeded to the unblinding only once a number of (previously defined) criteria where satisfied: essentially, we needed to have at least one set for each \g-ray map compatible with noise (because of K1) and one with some (possibly weak) signal (because of K2). This has been done by evaluating the $\chi^2$ differences defined in the main text as the statistical estimator. In order to evaluate the statistical significance of the obtained $\Delta\chi^2$, we derived the distribution of various cases from multivariate Gaussian realizations.
In the following, we summarise the main and most relevant results.

We note one substantial change to the analysis after unblinding. Tests with versions of the \Fermi~data that differ in the subtraction of a Galactic foreground model revealed a bug in our treatment of weak lensing shear around pixels with negative foreground-subtracted flux. All results in the main text of this paper have been updated, leading to an increase in signal-to-noise ratio from $4.5$ to $5.3$. The error did not affect our unblinding choices, particularly because it did not affect shear measurements around the G1 and G2 maps, which have no pixels with negative flux.

The \deltachi\ between null signal and models for all the combinations analyzed are reported in Table~\ref{tab:blindchi2}, for both the \pheno\ and \phys\ models. By looking at the \pheno\ model analysis, the table shows that the majority of cases have a very low \deltachi, as expected from the combination of the data sets discussed in the previous Sections. Specifically, for each of the versions of the blinded \g-ray maps, at least one of the three shear data vectors is consistent with pure shape noise. Two cases exhibit a somewhat large \deltachi: AZ and CX. There are two possibilities in the subsets data combination that could provide such a situation: the combination of any version of the \g-ray map with the \textsc{redMaGiC} galaxy signal or the combination of the true {\it Fermi} and the true DES data, if a signal is in fact present (combination G0-K0). These results, as well as those discussed below, remain valid also after the post-unblinding correction mentioned above.

\begin{figure}[t]
    \centering
    \includegraphics[width=0.45\textwidth]{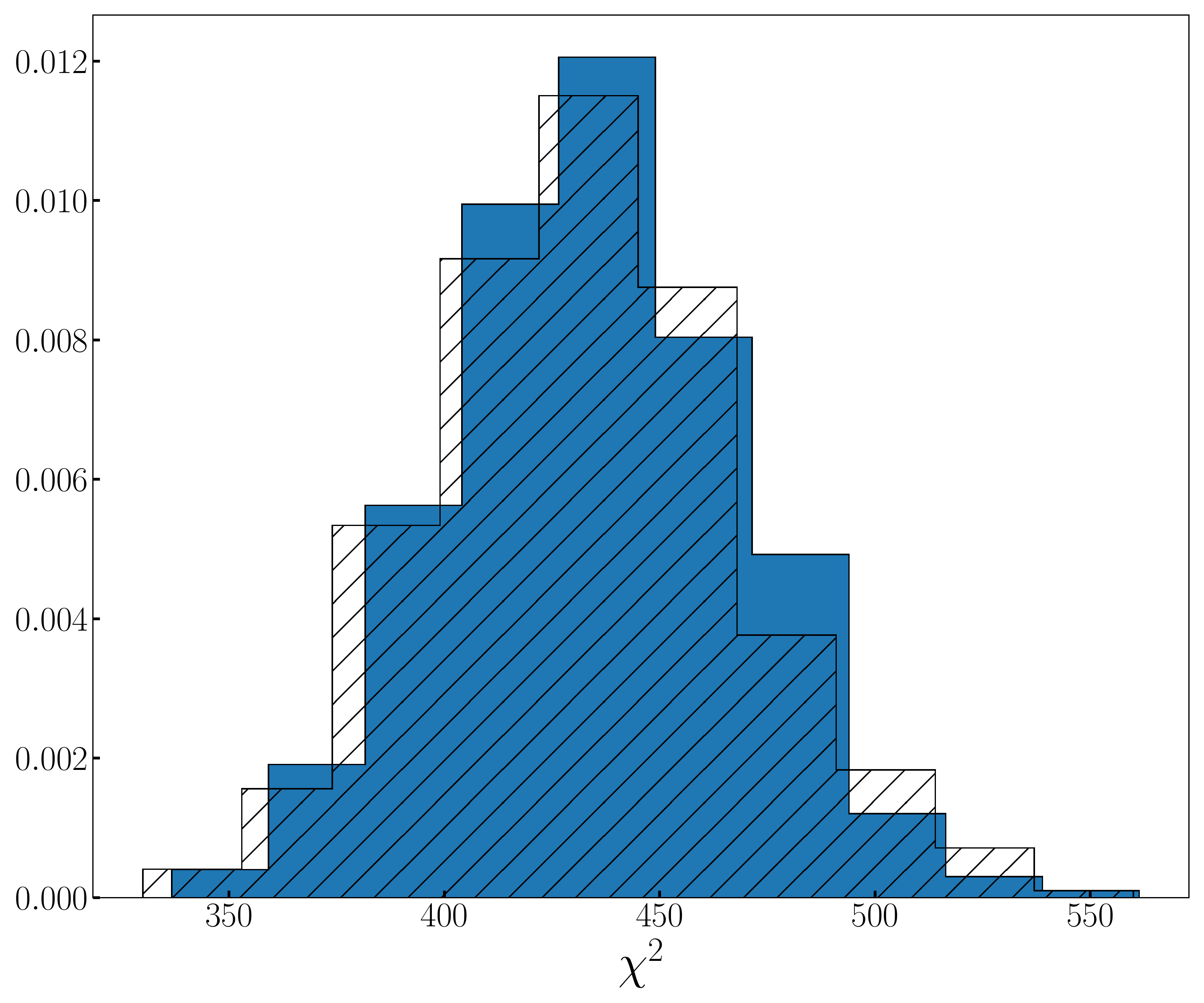}
    \includegraphics[width=0.45\textwidth]{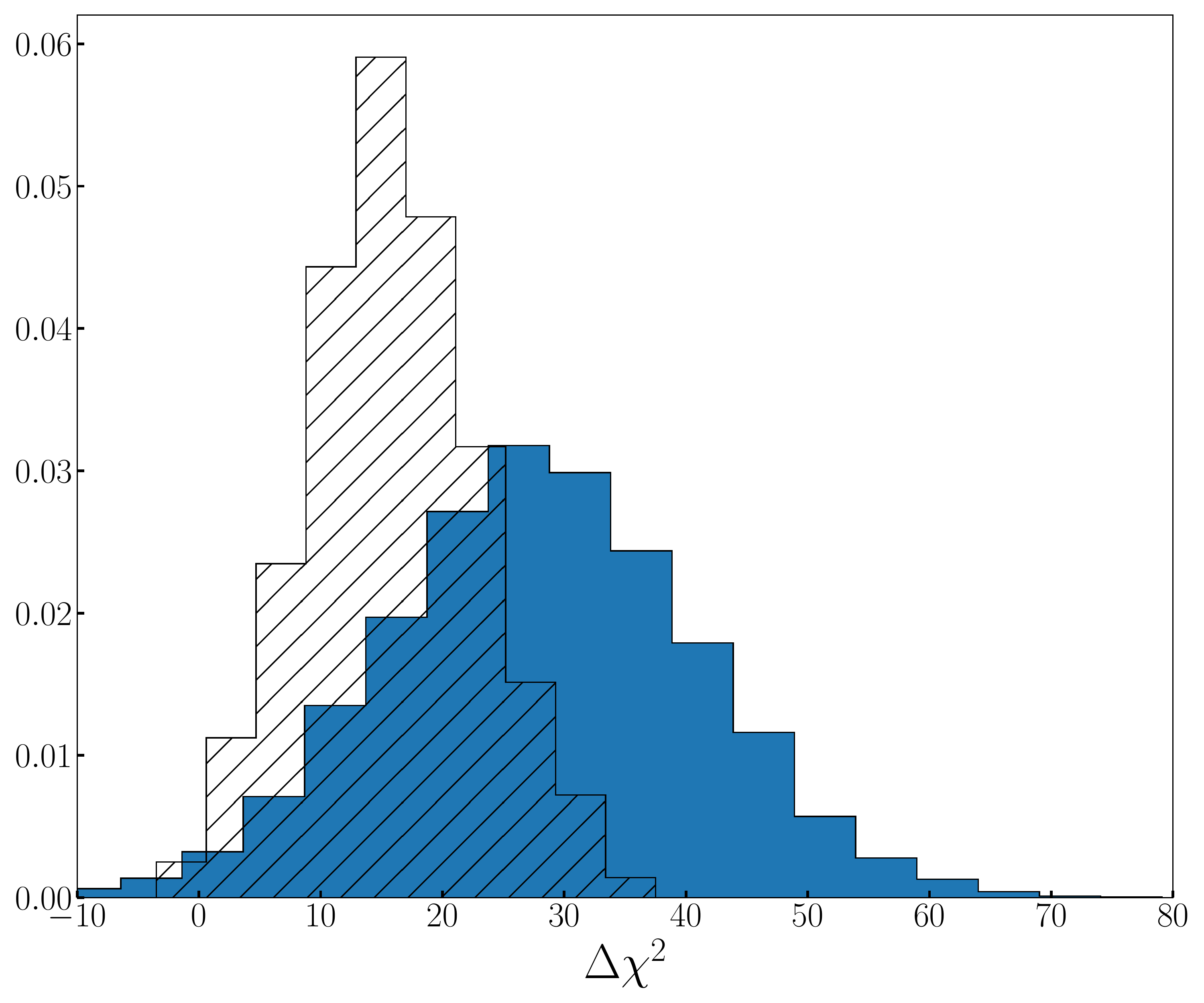}
    \caption{Distribution of the expected $\chi^2$ (left) and \deltachi\ (right) for the reference \phys\ model used in the Monte Carlo Markov Chain analysis during the blind phase (shaded areas) and the same distributions obtained after the post-unbliding correction mentioned above (blue areas). The normalization of the components in the model were chosen in order to reproduce the total UGRB emission. The meaning of the plot is to show the expectations for the null $\chi^2$ with the way of estimating the covariance used in this work, and for the \deltachi\ of the true case for the purpose of blinded tests. The \deltachi\ should be compared to the values reported in the second line of Table~\ref{tab:blindchi2}.}
    \label{fig:chi2distr}
\end{figure}

By looking at the \phys\ case results, we notice that the largest \deltachi's occur for CX, while AZ is consistent with no signal. The fact that AZ cannot be fitted by the \phys\ model implies this case does not provide a real signal. Let's comment that in this part of the analysis, for the \phys\ case we used a reference model (further discussed in the next Section) which assumes for simplicity the normalisation of the 1-halo and 2-halo terms for blazars to be equal ($A_{\rm BLZ}^{1{\rm h}}=A_{\rm BLZ}^{2{\rm h}}$) and which refers to a DM particle annihilating into hadronic states, specifically into a $\bar b b$ pair. The null $\chi^2$ distribution and the distribution of expected \deltachi\ for the reference \phys\ model are reported in Fig.~\ref{fig:chi2distr}. They are obtained by drawing from a multivariate Gaussian with mean given by the reference model and covariance given by the same covariance used in the data analysis. 
The null $\chi^2$ distribution peaks around 433.
The \deltachi\ distribution is rather broad and implies that \deltachi\ of all variants reported in Table~\ref{tab:blindchi2} are potentially compatible with the true case, since their \deltachi\ is well inside the distribution in Fig.~\ref{fig:chi2distr}. The peak of the distribution indicates that the expected \deltachi\ for the reference model is around 28. Fig. \ref{fig:chi2distr} shows as shaded areas the $\chi^2$ and $\Delta\chi^2$ distributions obtained before the post-unblinding correction was applied. No difference is obtained for the $\chi^2$, while for the $\Delta\chi^2$ a distribution peaked at larger values of $\Delta\chi^2$ is obtained after applying the correction, as a consequence of the increased sensitivity to the presence of a signal. The results are consistent with their mutual data sets and show that the bug did not lead to erronoeus conclusions.

\begin{table}
\centering
\begin{tabular}{cccccccccc}
 \hline
 & \multicolumn{9}{c}{Map combination} \\
 \cline{2-10}
 & AX & AY & AZ & BX & BY & BZ & CX & CY & CZ \\
 \hline
 \hline
$\Delta \chi^2_{\rm phe}$ & $1.17$ & $0.27$ & $29.3$ & $4.09$ & $0.26$ & $0.60$ & $27.5\,(21.1)$ & $1.56\,(1.27)$ & $1.50\,(1.90)$ \\
\hline
$\Delta \chi^2_{\rm phys}$ & $-0.06$ & $0.04$ & $2.58$ & $3.09$ & $-0.02$ & $3.33$ & $18.2\,(9.91)$ & $0.65\,(0.15)$ & $2.06\,(1.72)$ \\
 \hline
\end{tabular}
\caption{$\Delta\chi^2_{\rm mod}$ computed for the \phys\ and \pheno\ models with respect to the null hypothesis for the various combinations of the blind analysis. Only the combination of true \g-ray and shear data (CX) and the low-noise mock \g-ray map with an injected artificial shear signal (AZ) show a large preference for the model vs the null hypothesis (no signal). The {\it physical} model adopted in the blind analysis is the reference model discussed in the text. For CX, CY and CZ, we show in parenthesis also the values considered in the blinding phase, before the post-unblinding correction mentioned in the text.}
\label{tab:blindchi2}
\end{table}

The unblinding revealed the following identifications:
A $\xrightarrow{}$ G1, B $\xrightarrow{}$ G2, C $\xrightarrow{}$ G0, X $\xrightarrow{}$ K0, Y $\xrightarrow{}$ K1, Z $\xrightarrow{}$ K2. 
The coefficients of the linear combination in K2 were such that the injected signal was rather small, and therefore not easily identifiable. Moreover, being a signal due to the cross-correlation between the galaxy distribution and gravitational shear, it does not have to be compatible with our {\it physical} models (constructed specifically for the cross-correlation with \g-rays), while instead it could be well described by the {\it phenomenological} model (since it  contains a generic 1-halo and 2-halo terms).

From all these considerations, we see first of all that the results shown in Table~\ref{tab:blindchi2} are fully compatible with expectations: there is no ``spurious" detection, while the presence of a signal occurs only for those cases for which this is potentially possible. In fact, the CX case (the one corresponding to the combination of the true {\it Fermi} and DES data) is the only one that presents a high \deltachi\ for both the {\it phenomenological} and {\it physical} model. The \deltachi\ is larger in the case of the \pheno\ model since the latter has more freedom to adapt to data.
CY and CZ are compatible with null signal. The smallness of the injected signal in Z makes this case essentially indistinguishable from pure noise. Finally, the null $\chi^2$ for CX is 468, compatible with the expectations of the distribution of Fig. \ref{fig:chi2distr}.

None of the combinations involving A and B maps present a statistically significant \deltachi\ either with respect to the \phys\ or the \pheno\ model, except for the \pheno\ fit for the AZ case. We remind that map A is built from Poissonian \g-ray noise with enhanced photon count statistics. The latter means that the size of the errors is significantly reduced, and now the ``small'' injected signal is enough to provide a significant detection. On the other hand, the ``artificial'' nature of the signal is revealed by the fact that the \phys\ model is not able to fit it. The freedom we left to the \pheno\ model is instead large enough to make it able to include the signal of cross-correlation with galaxies.

In the next Section we outline the specifications of the reference model used for the blind analysis.

\section{Reference model}
\label{sec:refmodel}
\begin{figure*}[t]
    \centering
    \includegraphics[width=0.49\textwidth]{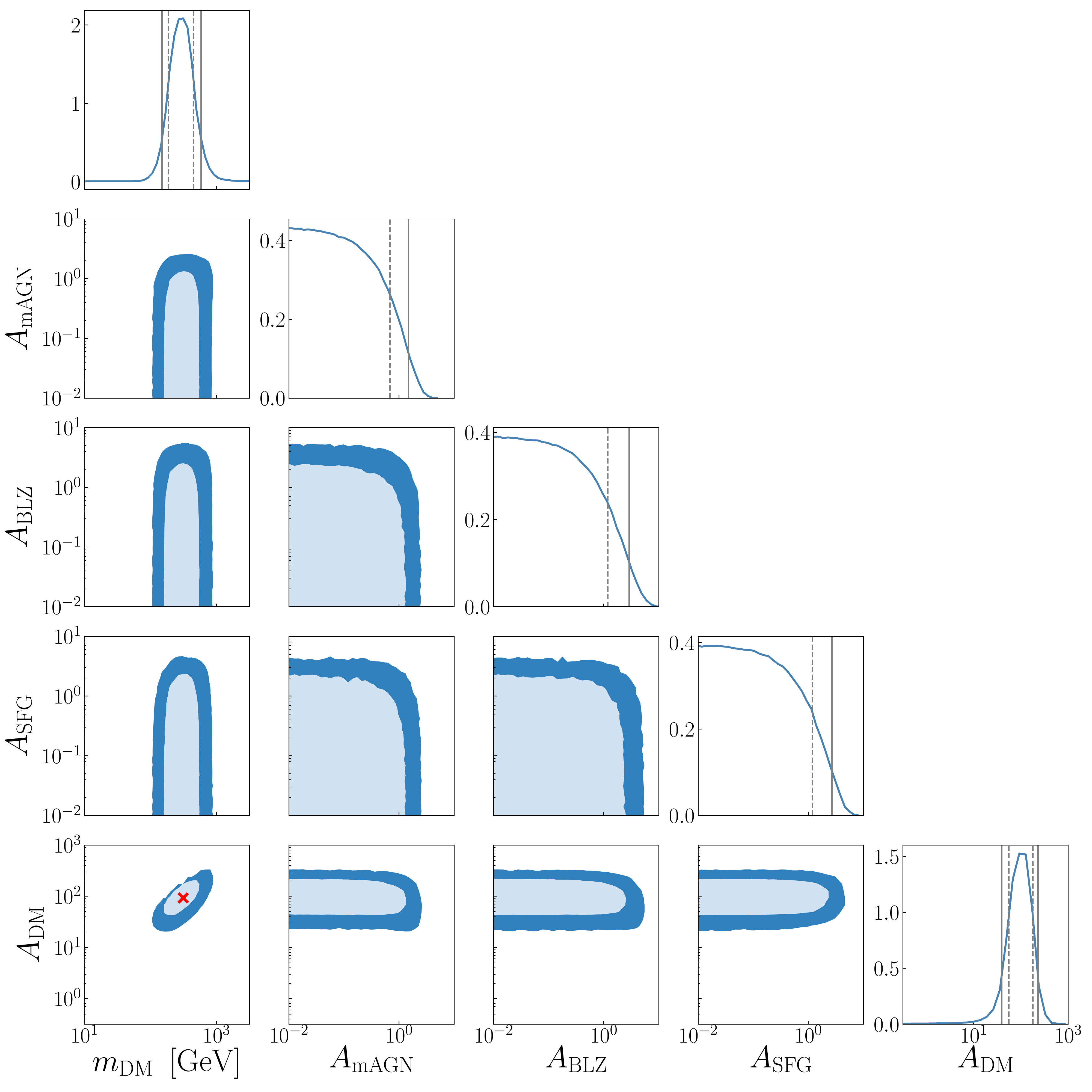}
    \includegraphics[width=0.49\textwidth]{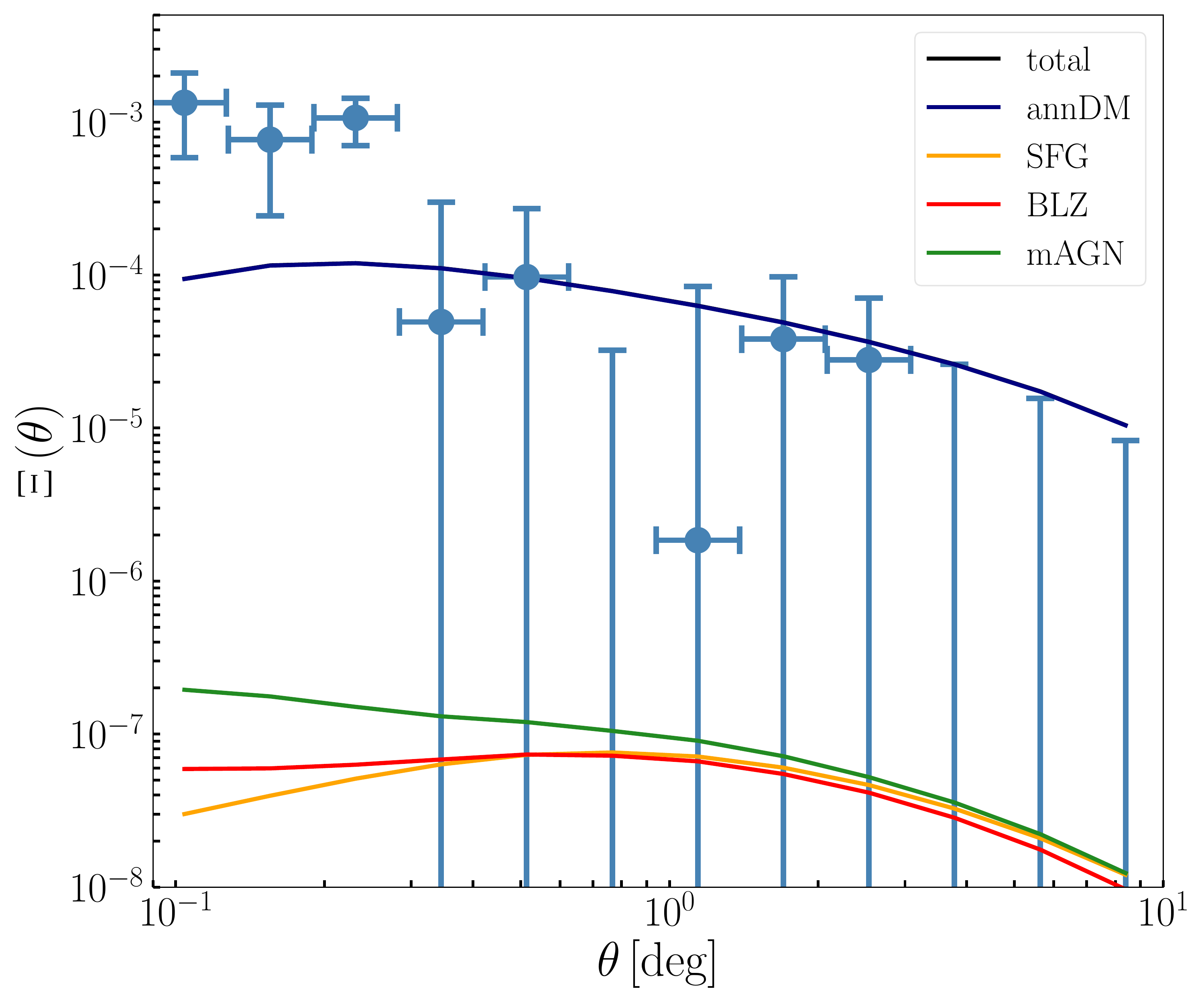}
    \caption{Left: Triangle plot for the normalisation parameters and \DM\ mass of the reference \phys\ model described in the text. The model incorporates star-forming galaxies, blazars, misaligned AGNs, and a \DM\ candidate annihilating in the $b\bar{b}$ channel. The parameters are presented in log-scale. The 1D profile likelihood distributions are normalised to unity. The dashed and solid lines represent, respectively, the 68\% and 95\% CL limits obtained in the Monte Carlo parameter scan. Same contour levels for the light and dark blue regions in the 2D distributions. Right: The blue dots show the integrated cross-correlation function obtained by averaging over all redshift and energy bins. The error bars are obtained from the diagonal terms of the covariance matrix, summed in quadrature.  Lines show the integrated best-fit CCF for the \phys\ {\it reference} model. The blue, orange, red and green lines correspond to the \DM\, star-forming galaxies, blazars, and misaligned AGNs contributions, respectively.  }
    \label{fig:bmrkmodel}
\end{figure*}
In the pipeline of the analysis before unblinding, we adopted a \phys\ model with blazars, mAGN, SFG and DM annihilating into a $\bar b b$ pair, with a common normalization for the 1-halo and 2-halo terms for blazars, i.e., $A_{\rm BLZ}^{1{\rm h}}=A_{\rm BLZ}^{2{\rm h}}$.
This was the first obvious options, since it adopts a nominal model for 
the blazar terms and minimises the number of free parameters.
As seen from the triangle plot in Fig.~\ref{fig:bmrkmodel}, the parameter scan for the CX case provides upper bounds for the three astrophysical components and a hint for the presence of a DM signal is found at the $3.1\sigma$ C.L., with best-fit parameters $m_{\rm DM} = 302$ GeV and normalization of the annihilation cross section relative to the natural scale $A_{\rm DM} = 105$.
Overall, the statistical significance of this reference \phys\ model is ${\rm SNR}=4.2$, with \deltachi$ = 18.2$ (for brevity, in this Section, we only quote the significance obtained after the post-unbliding correction mentioned above). 

In the right panel of Fig.~\ref{fig:bmrkmodel}, we show the cross-correlation signal for the best fit of this reference \phys\ model, compared to the data points.
While large scales are well fitted by the model, a clear lack of power is present at small angular scales. From Fig. 2 of the main text we instead know that in the \pheno\ model, the PSF-like 1-halo term can account for the measured CCF at small scales properly. This implies that the reference model, in its simplest formulation, is excessively limiting the ability to adapt to the data: a larger 1-halo term is needed.

Considering that the data require a hard-spectrum, not compatible with astrophysical components other than blazars, we improve the modelling by allowing the 1-halo and 2-halo terms for blazars to be independently normalised. In this case, the small scale correlation is well fitted, and a peak in the posterior distribution for the 1-halo blazar component arises (see Fig. 3 of the main text). This improves the ${\rm SNR}$ from $4.2$ to $5.2$, while the \DM\ component loses some of its significance with respect to the original reference case.

We have also investigated variations of the \DM\ model in terms of considering different annihilation channels, as it is usually done in \DM\ analyses. We found a preference for leptonic annihilation, with the preferred option being annihilation into $\tau$-lepton pairs: this increases the global ${\rm SNR}$ to $5.2$, with a preference for the presence of a \DM\ signal at the $2.8\sigma$ confidence level. This is the model that we report in the main text. In the case the DM component is not included, we obtain $\Delta \chi^2 = 16.5$, as compared to $\Delta \chi^2 = 27$ for the case with DM.

\begin{figure*}[t]
    \centering
    \includegraphics[width=0.45\textwidth]{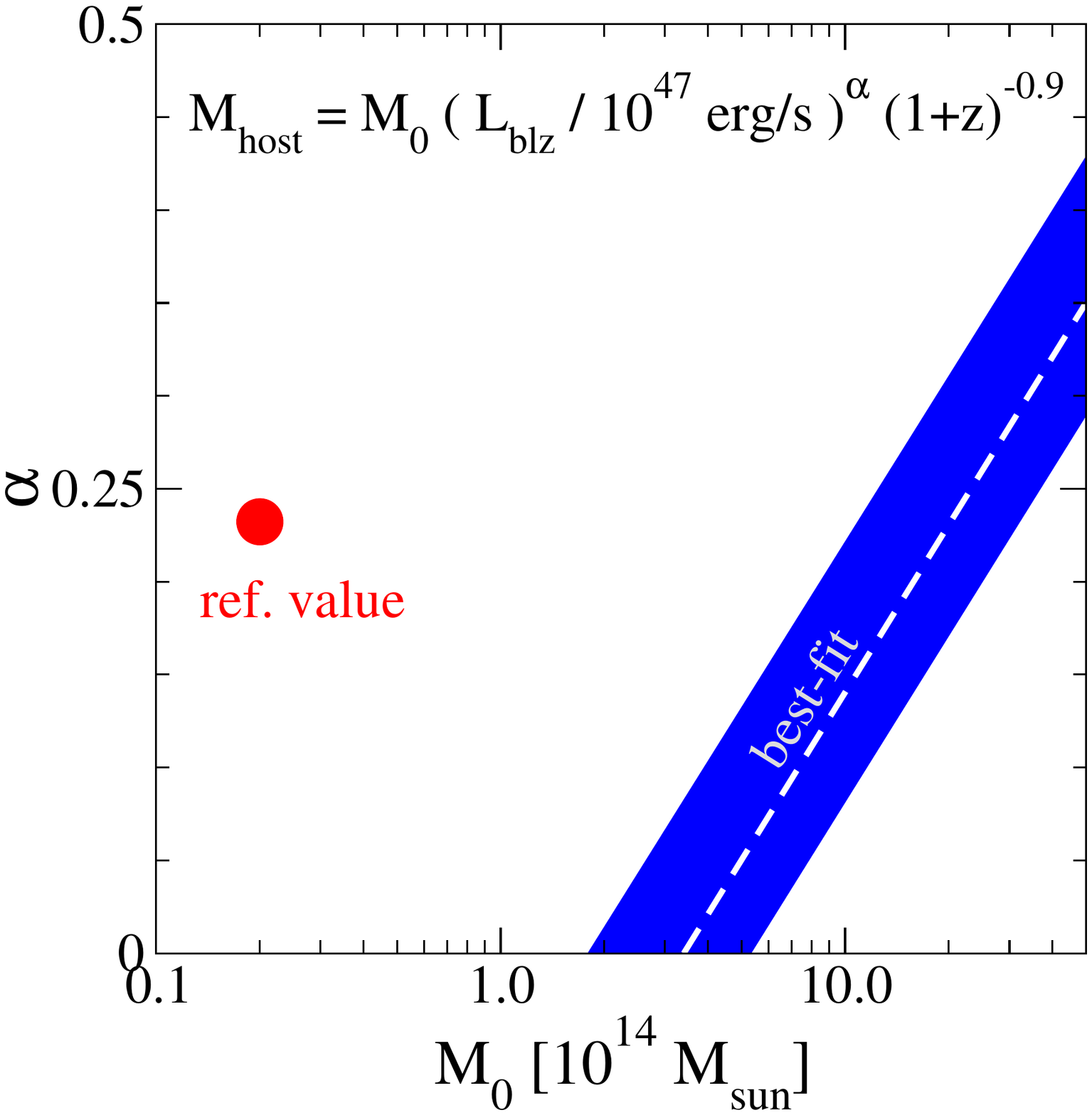}
    \includegraphics[width=0.5\textwidth]{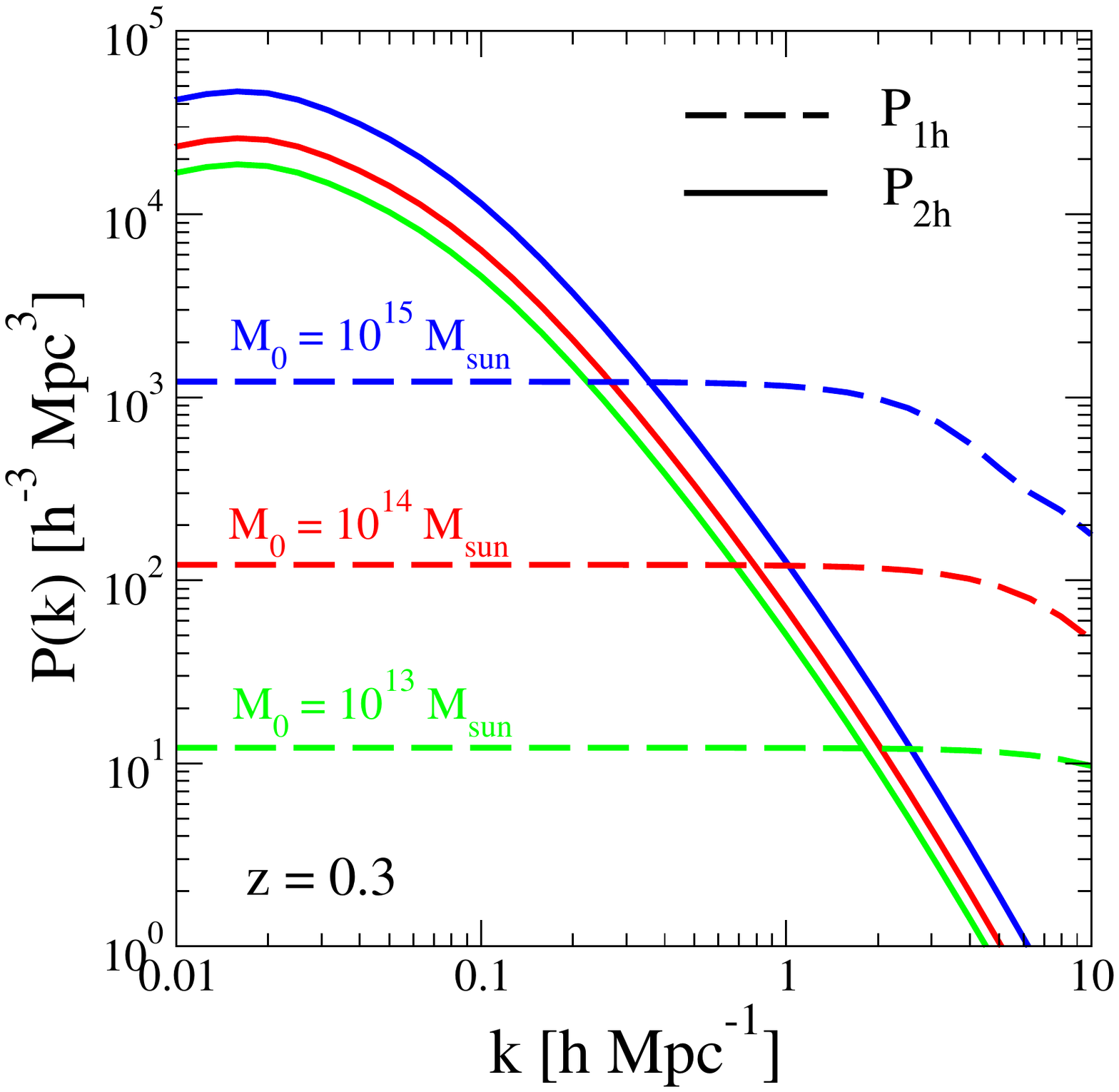}
    \caption{Left: The blue region shows the 68\% C.L. interval for the parameters $\alpha$ and $M_0$ describing the blazar luminosity versus host-halo mass relation. Red dot reports the combination adopted for the reference model described in the text. Right: Power spectrum of cross-correlation between \g-ray emission from blazars and shear at $z=0.3$ for different choices of the $M(\mathcal{L})$ relation. We show separately the 1-halo (dashed) and 2-halo terms (solid). For definiteness, we keep $\alpha$ fixed to 0.23 and vary only $M_0$.}
    \label{fig:DMsub2}
\end{figure*}

Concerning the blazar 1-halo component, this 
depends on the relation between the blazar luminosity and the host-halo mass, as can be seen in Eq.~\ref{eq:PSastro1h}. In the reference model used for the blind analysis 
this relation has been taken from ref. \cite{Camera:2014rja}, where it was derived by linking the \g-ray luminosity of a blazar to the mass of the supermassive black hole powering the AGN and then relating the mass of the black hole to the mass of the DM halo. Combining these pieces together, this peocedure gives $M(\mathcal{L})=2\times 10^{13}M_\odot\,(\mathcal{L}/10^{47}\mathrm{erg\,s^{-1}})^{0.23}(1+z)^{-0.9}$, where $\mathcal{L}$ is the rest-frame luminosity of blazars in the energy range $0.1$ to $100\,\mathrm{GeV}$.
Assuming this relation to be a power-law and fixing the redshift dependence as in the reference model (since, as we already mentioned, it is weakly constrained by data, and we therefore do not gain meaningful insight from the fit), we explore in the left panel of Fig.~\ref{fig:DMsub2} which $M(\mathcal{L})$ relation would be needed in order to reproduce the 1-halo term of our refined model. The plot shows the best-fit relation and its 68\% C.L.\ contours in the plane $(\alpha,M_0)$ for a relation $M(\mathcal{L})=M_0\,(\mathcal{L}/10^{47}\mathrm{erg\,s^{-1}})^\alpha (1+z)^{-0.9}$. Our results suggest the average mass of a halo hosting a blazar is larger than the one adopted in \cite{Camera:2014rja}, and most likely above $10^{14}M_\odot$. The cross-correlation signal with weak lensing seems therefore to be dominated by blazars residing in cluster-size halos. In the right panel of Fig.~\ref{fig:DMsub2}, we show with a few examples that modifying the $M(\mathcal{L})$ relation has a dramatic impact on the the 1-halo power spectrum, while it only mildly affects the 2-halo term.

\end{document}